\documentclass[lettersize,journal]{IEEEtran}
\usepackage{amsfonts,amsmath,amssymb,amsthm}
\usepackage[ruled,lined,linesnumbered]{algorithm2e}
\usepackage{amsbsy,caption}
\usepackage[para]{threeparttable}
\usepackage{graphicx,multirow,bm}
\usepackage{fancybox}
\usepackage{scalefnt}
\usepackage[noadjust]{cite}
\usepackage{tikz}
\usepackage{hyperref}
\usepackage{color}

\allowdisplaybreaks[4]

\newtheorem{Theorem}{Theorem}

\newtheorem{Example}{Example}

\newtheorem{Remark}{Remark}
\newtheorem{Lemma}{Lemma}

\newtheorem{Construction}{Construction}

\newcommand{\F}{\mathbb{F}}

\usepackage{pgf}
\usepackage{tikz}
\usetikzlibrary{matrix,chains,positioning,arrows,calc,decorations,plotmarks,patterns,trees,fit,backgrounds,shapes.multipart,topaths}
\usepackage{pgfplots}
\usetikzlibrary{external}                       % load externalization library
\usepackage{scalefnt}
\usepackage{tkz-graph}
\pgfplotsset{compat=1.3}
\tikzstyle{help lines}=[black!20,dashed]

\pgfplotscreateplotcyclelist{mylist}{red,blue,black,yellow,brown}

\pgfdeclarepatternformonly{my north east lines}{\pgfqpoint{-1pt}{-1pt}}{\pgfqpoint{8pt}{8pt}}{\pgfqpoint{6pt}{6pt}}%
{
	\pgfsetlinewidth{0.4pt}
	\pgfpathmoveto{\pgfqpoint{0pt}{0pt}}
	\pgfpathlineto{\pgfqpoint{6.1pt}{6.1pt}}
	\pgfusepath{stroke}
}
\definecolor{light_gray}{rgb}{0.6,0.6,0.6}
\definecolor{awgray}{rgb}{0.7,0.7,0.7}
\definecolor{awgray_dark}{rgb} {0.4,0.4,0.4}
\tikzset{
	>=stealth',
	mycircle/.style={circle, draw=gray, very thick},
	mycircle_small/.style={circle,draw=awgray_dark,fill = awgray_dark, inner sep=0,minimum size=.6em},
	mycircle_small_black/.style={circle,draw=black,fill = black, inner sep=0,minimum size=.6em},
	mybox/.style={rectangle,rounded corners,draw=black, thick,text width=1em,minimum height=4em,minimum width=4em,text centered},
	mybox_small/.style={rectangle,rounded corners,draw=black, thick,text width=1em,minimum height=2em,minimum width=2em,text centered},
	mybox_big/.style={rectangle,rounded corners,draw=black, thick,text width=12.5em,minimum height=7.8em,minimum width=12.5em,text centered},
	mybox_vec/.style={rectangle,rounded corners,draw=black, thick,text width=10em,minimum height=5em, minimum width=10em,text centered},
	mybox_vec_short/.style={rectangle,rounded corners,draw=black, thick,text width=1em,minimum height=0.7em, minimum width=2em,text centered},
	pil/.style={->, thick, shorten <=2pt, shorten >=2pt,},
}

\begin{document}
	
	\title{A New Cooperative Repair Scheme with Small Finite Field for Distributed Storage Systems}
	
	\author{Yajuan Liu,~\IEEEmembership{Member,~IEEE},  Han Cai,~\IEEEmembership{Member,~IEEE}, and Xiaohu Tang, \IEEEmembership{Fellow, IEEE}
		\thanks{Y. Liu is with the Electrical and Electronics Engineering Department, Bilkent University, Ankara, Turkey (email: yajuan.liu@bilkent.edu.tr), and H. Cai, and X. Tang are with the Information Security and National Computing Grid Laboratory, Southwest Jiaotong University, Chengdu, China (email: hancai@swjtu.edu.cn, \mbox{xhutang@swjtu.edu.cn}).
		}
		\thanks{The material in this paper was presented in part in the IEEE 23rd International Conference on Communication Technology 2023.}
		%\thanks{This work was supported in part by the National Natural Science Foundation of China under Grant 62271421.}
	}

	\maketitle
	
	\begin{abstract}
		In this paper, we consider the multiple failures in the distributed storage systems under the cooperative repair model.
		We introduce a new cooperative repair scheme for the $(n,k,d,N)$ minimum storage regenerating (MSR) codes proposed by Ye and Barg (IEEE Transactions on Information Theory, vol. 64, no. 4, 2017), which is capable of repairing any $h$ failed nodes by connecting any $k\le d\le n-h$ helper nodes.
		%	To the best of our knowledge, this is the first optimal cooperative repair scheme specifically for the MSR codes with optimal access and update properties.
		Compared to prior cooperative repair schemes for $(n,k,d,N)$ MSR codes, which require a finite field $\mathbb{F}_q$ with $q\ge (d-k+1)n$, the proposed approach reduces the field size to $q\ge n+1$.
		%reduces  the size of finite field to $q \ge n + 1$.
	\end{abstract}
	
	\begin{IEEEkeywords}
		Distributed storage systems, optimal repair bandwidth, cooperative repair, MSR codes
	\end{IEEEkeywords}
	
	\section{Introduction}
	Nowadays, large-scale distributed storage systems such as the Google File System (GoogleFS) and Amazon Elastic Block Store (EBS) have become fundamental infrastructure in modern computing \cite{ramkumar2022codes}. In these systems, coding technologies have played an essential role in protecting data against node erasures. 
	Specifically, an $(n,k)$ code encodes $k$ information
	symbols into $n$ coded symbols distributed over $n$ storage nodes. 
	Among all possible $(n,k)$ codes, those achieving the largest possible Hamming distance, namely, maximum distance separable (MDS) codes, have attracted significant attention in distributed storage systems, since a larger Hamming distance provides stronger resilience against node erasures. 
	However, from the perspective of node repair, MDS codes suffer from a well-known drawback. When repairing a single failed node using $k$ helper nodes, the system must download all the data stored in these $k$ nodes, resulting in a repair bandwidth equal to the size of the original file, even though only one node is lost.
	
	To improve the repair efficiency of storage codes, regenerating codes were introduced in \cite{dimakis2010network}, which first regard the repair bandwidth as an important parameter for codes and optimize the trade-off between storage and repair bandwidth. Among regenerating codes, minimum storage regenerating (MSR) codes are of particular interest, as they possess both the MDS property and optimal repair bandwidth.
	
	In recent years, significant efforts have been devoted to the construction of MSR codes, especially for the single failure scenario \cite{balaji2018erasure,chen2019explicit,hou2019rack,li2021systematic,papailiopoulos2013repair,shao2017tradeoff,tamo2011zigzag,wang2012long,li2018generic,ye2017explicit,zhang2023vertical,wang2023rack,patra2025generalized,jiang2024generalized,li2024MSR,zeng2025an,li2024generalization}.
	However, in practical distributed storage systems, to avoid frequent scanning and repairs, the system may wait until the number of erasures reaches a certain threshold. Therefore, multiple failures should also be considered.
	In general, for repair mechanisms of multiple failures, two main models have been considered: centralized repair and cooperative repair.
	In the centralized repair model, data of failures is recreated in a data center and then transferred to new nodes.
	In the cooperative repair model, failed nodes collaboratively recover their data by downloading information from helper nodes and exchanging data among themselves, where the system does not require a centralized data center for repair.
	The latter is particularly attractive due to its distributed nature and improved bandwidth efficiency.

	Specifically, under the cooperative repair model of an $(n,k,N)$ MDS code, assume that $h$ failures occur.
	To repair them, the failed nodes connect to $k\le d\le n-h$ helper nodes.
	The repair process  is composed of two phases \cite{shum2013cooperative}:
	\begin{enumerate}
		\item \textbf{Download phase:} Each failed node downloads $\beta_1$ coded symbols from each of $d$ helper nodes,  respectively.
		
		\item \textbf{Cooperative  phase:} For any two failed nodes,  each transfers $\beta_2$ symbols to the other.
	\end{enumerate}
	Accordingly,  the repair bandwidth is $\gamma=h(d\beta_1+(h-1)\beta_2)$.
	
	In \cite{hu2010cooperative} and \cite{shum2013cooperative},
	the repair bandwidth of an MDS array code $\mathcal{C}$ is proved to be lower bounded by
	\begin{eqnarray}\label{eqn:Lower_Bound}
		\gamma\ge \frac{h(d+h-1)N}{d-k+h},
	\end{eqnarray}
	where $N$ is the length of data stored in each node, called as \textit{sub-packetization level}.
	Especially, the optimal repair bandwidth $h(d+h-1)N/(d-k+h)$ in \eqref{eqn:Lower_Bound} is achieved only for $\beta_1= \beta_2=N/(d-k+h)$ \cite{shum2013cooperative}.
	In this case, the MDS array code $\mathcal{C}$ is known as \textit{$(n,k,d,h,N)$ MSR code}.
	
	In the literature, there are three well-known MSR codes: (i) Diagonal matrix-based codes, such as Hadamard MSR codes \cite{papailiopoulos2013repair} and construction $1$ in \cite{ye2017explicit}, (ii) Hybrid matrix-based codes, like Long MDS codes \cite{wang2012long} and those presented in \cite{li2024MSR}, and (iii) Permutation matrix-based codes, such as Permutation codes \cite{cadanbe2011permutation}, Zigzag MSR codes \cite{tamo2011zigzag} and Construction $4$ in \cite{ye2017explicit}. Generally, from a practical perspective, the sub-packetization $N$ of MSR codes determines the encoding and decoding complexity, and the size of finite field $q$ has a major impact on the implementation complexity.
	Thus, it is desirable to construct MSR codes with small sub-packetization $N$ and field size $q$.
	
	\begin{table*}[!ht]
		\centering
		\begin{center}\caption{Comparisons of $(n,k,d,h,N)$ MSR codes with optimal cooperative repair property.}\label{comparison}
			\renewcommand\arraystretch{1.2}
			\begin{tabular}{|c|c|c|c|c|}
				\hline
				MSR codes &	Sub-packetization $N$   & Helper nodes $d$ & Finite field $\mathbb{F}_q$  & Ref.\\
				\hline
				\multirow{4}{*}{Diagonal matrix-based}&	$\left((d-k)^{h-1}(d-k+h)\right)^{n \choose h}$  & $k\le d \le n-h$  &$(d-k+1)n$ & Ye and Barg \cite{ye2018cooperative}\\	
				\cline{2-5}
				&	$(d-k+h)(d-k+1)^n$& $k\le d \le n-h$ &$(d-k+1)n$ & Ye \cite{ye2020new}\\
				\cline{2-5}
				&	$\left\{\begin{array}{ll}
					2^n, & \mathrm{if}~(h+1)|2^n\\
					(2\ell+1)2^n, & \mathrm{if}~ h+1=(2\ell+1)2^m,  \ell\ge 1
				\end{array}
				\right.$   &  $d=k+1$&$(d-k+1)n$ &Liu \textit{et al.} \cite{liu2022new} \\
				\hline
				\multirow{2}{*}{Hybrid matrix-based}&	$(d-k+h)^{n \choose h}$ & $k\le d \le n-h$ & $n+d-k$ & Zhang \textit{et al.} \cite{zhang2020explicit}\\
				\cline{2-5}
				&	$(d-k+h)(d-k+1)^{\lceil{n/2}\rceil}$& $k\le d \le n-h$ &$(d-k+1)n+1$ & Zhang \textit{et al.} \cite{zhang2024construction}\\
				\hline
				Permutation matrix-based&	$(d-k+h)(d-k+1)^n$& $k\le d\le n-h$ &$n + 1$& This paper \\
				\hline
			\end{tabular}
		\end{center}
	\end{table*}
	
	For MSR codes under cooperative repair, existing works have primarily focused on reducing the sub-packetization level $N$ while maintaining optimal repair bandwidth.
	Focus on the diagonal matrix-based MSR codes,
	Ye and Barg constructed the optimal cooperative repair scheme with the sub-packetization level $N=\left((d-k)^{h-1}(d-k+h)\right)^{n \choose h}$ in \cite{ye2018cooperative}.
	Subsequently,
	Ye   dramatically reduced the sub-packetization  to $(d-k+h)(d-k+1)^n$ in \cite{ye2020new}.
	Recently,  Liu \textit{et al.} \cite{liu2022new} removed or reduced the multiplication coefficients (i.e., $d-k+h$)  for $d=k+1$ further.
	With regard to the hybrid matrix-based MSR codes, Zhang \textit{et al.}  proposed the cooperative repair scheme in \cite{zhang2020explicit} with the sub-packetization  $(d-k+h)^{n \choose h}$. Very recently, Zhang  \textit{et al.} \cite{zhang2024construction}  significantly  lowered the sub-packetization  to $(d-k+h)(d-k+1)^{\lceil n/2\rceil}$.

	However, these constructions typically require relatively large finite field sizes (e.g., $\mathbb{F}_q$ with $q\ge (d-k+1)n$), which increases computational complexity and limits practical applicability. As shown in \cite{pedersen2008implementation} and \cite{vingelmann2009impletation}, the impact of the field size on implementation efficiency can be substantial, demonstrating the practical benefits of operating over smaller finite fields. In addition, modern large-scale storage systems also require flexibility in supporting a wide range of code parameters under a fixed implementation platform, as reflected in recent industrial storage architectures \cite{white2025VAST,breaking2019VAST}.
	
	Among known MSR codes, the permutation matrix-based construction in \cite{ye2017explicit} achieves one of the smallest known field sizes and supports a broad range of admissible code lengths under the same small-field setting. This combination is particularly attractive in practice, as it simultaneously addresses implementation efficiency and parameter flexibility. Moreover, these codes achieve the optimal access property for single-node repair, thereby reducing I/O overhead. Nevertheless, existing studies on this construction mainly focus on single-node repair, and cooperative repair for such codes remains largely unexplored.

	This paper addresses this gap by developing a general cooperative repair framework for permutation matrix-based MSR codes in \cite{ye2017explicit}.
	In contrast to existing works, which either focus on different code structures or require large field sizes, the proposed scheme achieves optimal cooperative repair bandwidth while significantly reducing the required finite field size to $q\ge n+1$, which is the smallest known among comparable constructions.
	The key idea is to combine space sharing with a structured summation and grouping strategy, so that interference can be aligned precisely and the resulting coefficient matrix remains invertible.
	Importantly, the proposed scheme is not a direct extension of existing cooperative repair methods. Instead, it is a dedicated design that exploits the algebraic structure induced by permutation matrices. % Our work provides the 
	Specifically, our approach involves two key steps: First, we apply space sharing technique to create $d-k+h$ instances of the original MSR codes, then elaborately sum $d-k+1$ instances to generate some new parity check equations. Next, these equations are partitioned into subgroups, each containing $r=n-k$ equations. By systematically combining $(d-k+1)^{h-1}$ subgroups, our scheme ensures that the number of unknowns precisely matches the number of equations and  the coefficient matrix of the unknowns in the parity check equations of those groups is invertible, which achieves perfect interference alignment, thereby guaranteeing the recovery of the failed nodes.
	Table \ref{comparison}
	compares the proposed scheme with some existing works.

	To the best of our knowledge, this is the first work that systematically investigates optimal cooperative repair schemes for permutation matrix-based MSR codes with small finite field size.
	The main contributions of this paper can be summarized as follows:
	
	i) We propose a general cooperative repair scheme for permutation matrix-based MSR codes in \cite{ye2017explicit} that supports arbitrary numbers of failed nodes, flexible helper node selection, and a wide range of code lengths. The proposed scheme achieves the optimal cooperative repair bandwidth while significantly reducing the required finite field size to $q \ge n+1$.
	
	ii) We provide a systematic construction and analysis framework, including the design of parity check equations and the proof of the invertibility of the recovery matrix.

	The rest of the paper is organized as follows. Section \ref{sec:preliminary} reviews some necessary preliminaries. Section \ref{sec:repair scheme} presents the cooperative repair scheme of any $h$ failed nodes, while Section \ref{sec:h=2,3}  analyzes its correctness and performance. In Section \ref{sec:conclusion}, we conclude this paper.

	\section{Preliminaries}\label{sec:preliminary}
	In this section, we briefly review the permutation matrix-based $(n,k,d,h,N)$ MSR codes proposed in \cite{ye2017explicit}.
	For ease of reading,  we summarize some useful notations in Table \ref{tab:notations}.
	\begin{table*}
		\caption{Main notations in this paper.}\label{tab:notations}
		\centering	\begin{tabular}{|l|l|}
			\hline
			\multirow{2}{*}{$(n,k,d,h,N)$} & an MDS code with codelength $n$, dimension $k$, and sub-packetization $N$.\\
			& When $h$ nodes fail, one connects to $d$ helper nodes to repair the failures\\
			\hline
			$s$ and $\mathbb Z$ & the parameter $s=d-k+1$ and the ring of integers\\
			\hline
			$\mathbf{f}\in\mathbb{F}_q^N$ & a vector of length $N$ over the finite field $\mathbb{F}_q$, where $q$ is a prime power\\
			\hline
			$[a,b)$ and $[a,b]$ & $\{a,a+1,\dots,b-1\}$ and $\{a,a+1,\dots,b\}$\\
			\hline
			$\mathbf{e}_{N,a}, a\in[0,N)$ & a  binary vector of length $N$ with only the $a$-th component being non-zero\\
			\hline
			$\mathbf{a}=(a_0,a_1\dots,a_{n-1})$  & the $s$-ary representation  of an integer $a\in[0,s^n)$, i.e., $a=\sum_{i=0}^{n-1}a_i\cdot s^i,a_i\in[0,s)$ \\
			\hline
			$a(i,\pm j)$ & $(a_0,\dots,\langle a_i\pm j\rangle,\dots,a_{\bar{n}-1})$, where  $\langle \cdot \rangle$ means the smallest nonnegative residue of $\cdot$ modulo $s$\\
			\hline
			$a(i_1,b_1;i_2,b_2)$ & $(a_0,\dots,\langle a_{i_1}+b_1\rangle,\dots,\langle a_{i_2}+b_2\rangle,\dots,a_{\bar{n}-1})$\\
			\hline
			$A(i)$  and $A(i,j)_{m\times n}$
			& the $i$-th row and the entry in the $i$-th row and $j$-th column of an $m\times n$ matrix $A$\\
			\hline
			$\zeta_{t,i,a}$ & the coefficient of the $a$-th symbols of the $t$-th parity matrix of node $i$\\
			\hline
			$\mathcal{A}_{\ell},\mathcal{B}_{\ell,\tau}$& the sets of symbols in the $\ell$-th group and the $(\ell,\tau)$-th subgroups of parity check equations\\
			\hline
			$\mathcal{F}=\{i_0,\dots,i_{h-1}\}$ & the set of $h$ failed nodes\\
			\hline
			$\mathcal{H}=\{j_0,\dots,j_{d-1}\}\subseteq[0,n)\backslash \mathcal{F}$ & the set of $d$ helper nodes\\
			\hline
			$\mathcal{U}=[0,n)\backslash(\mathcal{F}\cup \mathcal{H})=\{i_h,\dots,i_{n-d-1}\}$ & the set of $n-d-h$ unconnected nodes \\
			\hline
		\end{tabular}
	\end{table*}

	\subsection{Permutation matrix-based MSR codes}\label{subsection:zigzag}	
	For an $(n,k)$ MDS code, if the data stored in each \mbox{node} is a column vector of length $N$ over the finite field $\mathbb{F}_q$ rather than a scalar, we refer to it as \textit{an $(n,k,N)$ MDS array code}.
	The \textit{$(n,k,d,h,N)$ MSR code} is a class of $(n,k,N)$ MDS array codes with the optimal repair bandwidth.
	Assume that the original data is of size $M=kN$.
	An $(n,k,d,h,N)$ MSR code partitions the data into $k$ parts and then encodes them to $n$ parts $\mathbf{f}=[\mathbf{f}_0^\top,\dots,\mathbf{f}_{n-1}^\top]^\top$  stored across $n$ nodes,
	where $\mathbf{f}_i=(f_{i,0},\dots,f_{i,N-1})^\top\in \mathbb{F}_q^N,i\in [0,n)$ is a column vector and $\top$ denotes the transpose operator.	
	The $(n,k,d,h,N)$ MSR codes can be defined by the following parity check equations,
	\begin{IEEEeqnarray}{c}\label{Parity-check equation}
		A_{t,0}\mathbf{f}_0+ A_{t,1}\mathbf{f}_1+\cdots+ A_{t,n-1}\mathbf{f}_{n-1}=\mathbf{0},
	\end{IEEEeqnarray}
	where $A_{t,i},t\in[0,r=n-k),i\in[0,n)$ is an $N\times N$ nonsingular matrix over $\mathbb{F}_q$,
	called the \textit{parity matrix} of node $i$
	for the $t$-th parity check equation, and $\mathbf{0}$ denotes an all-zero column vector of length $N$.
	In matrix form, the structure of $(n,k,d,h,N)$ MSR codes based on the above parity check equations  can be rewritten as
	\begin{eqnarray*}\label{Matrix form}
		\underbrace{\left(\begin{array}{ccccc}
				A_{0,0} & A_{0,1} & \cdots  &   A_{0,n-1}      \\
				A_{1,0} & A_{1,1} & \cdots  &   A_{1,n-1}      \\
				\vdots  & \vdots  &  \ddots  &  \vdots          \\
				A_{r-1,0} & A_{r-1,1} & \cdots  &   A_{r-1,n-1}      \\
			\end{array}\right)}_{\mathrm{block ~matrix~}  A}
		\left(\begin{array}{ccccc}
			\mathbf{f}_0        \\
			\mathbf{f}_1\\
			\vdots\\
			\mathbf{f}_{n-1}\\
		\end{array}
		\right)=\mathbf{0}.
	\end{eqnarray*}
	
	Usually, $A$ is designed as a block Vandermonde matrix, i.e.,
	\begin{eqnarray*}
		A_{t,i}=A_i^t,\qquad i\in[0,n), ~t\in[0,r),
	\end{eqnarray*}
	where $A_i,i\in[0,n)$ is an $N \times N$ nonsingular matrix.
	In particular, we use the convention $A_i^0=I$.

	%	To  illustrate $(n,k,d,h=1,N)$ Zigzag MSR codes, we firstly define some parameters here.
	
	To  illustrate $(n,k,d,h=1,N)$ permutation matrix-based MSR codes, we firstly define some parameters here.
	Let $N=s^n$ and $\gamma$ be a primitive element over $\mathbb{F}_q,q\ge n+1$, where  $s=d-k+1$ is an integer.
	For any $i\in[0,n)$,
	denote
	\begin{eqnarray}\label{zigzag:A_i:parameter}
		\lambda_{i,z}=\left\{\begin{array}{lll}
			\gamma^{i+1}, &\text{ if } z=0,\\
			1, & \text{ else.}
		\end{array}	
		\right.
	\end{eqnarray}
	Then, for any $\mathbf{a}=(a_0,\dots,a_{n-1})\in[0,s^n)$, set
	\begin{eqnarray}\label{eqn:zeta}
		\zeta_{t,i,a} = \left\{\begin{array}{lll}
			1, &\text{ if } t=0,\\
			\prod\limits_{0\leq m < t}\lambda_{i,\langle a_i+m\rangle}, & \text{ else.}
		\end{array}	
		\right.
	\end{eqnarray}
	
	Denote the $a$-th row vector of $A_{t,i},t\in[0,r),i\in[0,n),a\in[0,s^n)$ by
	\begin{equation}\label{eqn:A_{t,i,a}}
		A_{t,i}(a)=\zeta_{t,i,a}\mathbf{e}_{N,a(i,t)}.
	\end{equation}
	Note that $A_{t,i}$ is a permutation matrix that moves the $a(i,t)$-th row to the $a$-th row, multiplying the row by $\zeta_{t,i,a}$.
	
	Thereafter, an $(n,k,d,h=1,N)$  MSR  code $\mathcal{C}'$ with the sub-packetization $N=s^n$ can be characterized by the parity matrices below,
	\begin{eqnarray}\label{A_i^t}
		A_{t,i}&=&
		\left(A_{t,i}(0)^\top,\dots,A_{t,i}(N-1)^\top\right)^\top \nonumber\\
		&=&
		\sum_{a=0}^{N-1}\zeta_{t,i,a}\mathbf{e}_{N,a}^\top \mathbf{e}_{N,a(i,t)},\quad i\in[0,n),t\in[0,r).
	\end{eqnarray}
	
	Definitely, according to \eqref{zigzag:A_i:parameter} and \eqref{eqn:zeta}, we have
	\begin{eqnarray*}\label{eqn:A_1,i}
		A_{1,i}=A_i= \sum\limits_{a=0}^{N-1}\lambda_{i,a_i}\mathbf{e}_{N,a}^\top \mathbf{e}_{N,a(i,1)}  ,\quad i\in[0,n),
	\end{eqnarray*}
	and
	\begin{eqnarray*}
		A_{t,i}=A_i^t=\gamma^{(i+1)z} I_{s^n},\quad i\in[0,n),t = zs,z\in \mathbb{Z},
	\end{eqnarray*}
	where $I_{s^n}$ is an $s^n\times s^n$ identity matrix.

	%\begin{Remark}
	%The parity check matrix of $(n,k,d,h=1,N)$  MSR  code $\mathcal{C}'$ can be represented using the Kronecker product as follows:
	%\begin{eqnarray*}\label{Matrix form}
	%		A=\left(\begin{array}{ccccc}
		%				A_{0,0} & A_{0,1} & \cdots  &   A_{0,n-1}      \\
		%				A_{1,0} & A_{1,1} & \cdots  &   A_{1,n-1}      \\
		%				\vdots  & \vdots  &  \ddots  &  \vdots          \\
		%				A_{r-1,0} & A_{r-1,1} & \cdots  &   A_{r-1,n-1}      \\
		%			\end{array}\right)
	%	\end{eqnarray*}
%and
%$$
%A_{t,i} = I_{s^{n-i-1}} \otimes (E_{t,i} \otimes I_{s^i}), \quad i \in [0,n),\ t \in [0,r),
%$$
%where $\otimes$ denotes the Kronecker product, $I_m$ is the $m \times m$ identity matrix, and
%$E_{t,i},t\in[0,r),i\in[0,n)$ is an $s\times s$ matrix with
%		\begin{eqnarray*}
	%			E_{t,i}(a,j)=\left\{\begin{array}{lll}
		%				\zeta_{t,i,a}, &\text{ if } j=\langle a+t\rangle,\\
		%				0, & \text{~else.}
		%			\end{array}\right.
	%		\end{eqnarray*}
%	
%This representation is more compact than \eqref{A_i^t}. However, since we need to explicitly determine the coefficient of each coordinate at each node in this paper, we opt to characterize each entry of the parity-check matrix to facilitate a detailed analysis of the parity-check equations.
%\end{Remark}

Additionally, we can get the following two properties.
\begin{itemize}
	\item [P1.]
	For any $t\in[0,r), i \in[0,n),a\ne b\in[0,s^n), \zeta_{t,i,a}=\zeta_{t,i,b}$ if $a_i=b_i$.	
	
	\item [P2.]
	For any $0\le i\ne j<n$,
	$\zeta_{t,j,a(i,w)}= \zeta_{t,j,a},a\in[0,s^n),w\in[0,s)$.
	
\end{itemize}

Subsequently, we reveal the $a$-th row of the $t$-th parity check equation below,
\begin{eqnarray}\label{eqn:a-row}
	\sum_{i=0}^{n-1}A_{t,i}(a)\mathbf{f}_i
	=
	\sum_{i=0}^{n-1}\zeta_{t,i,a}f_{i,a(i,t)} = 0,\nonumber\\
	a\in[0,s^n),t\in[0,r).
\end{eqnarray}

	\section{The General Repair Strategy of MSR Codes}\label{sec:repair scheme}
	
	In this section, we propose a general repair scheme for the multiple failures of permutation matrix-based $(n,k,d,h,N)$  MSR codes with $N= (d-k+h)(d-k+1)^n$.
	We first construct the desired MSR codes via space sharing multiple instances of the original permutation matrix-based MSR codes, and then develop a tailored cooperative repair scheme for $h$ failed nodes by carefully organizing the parity check equations through summation and grouping strategies.
	Throughout this paper, it is assumed that $h$ nodes of an $(n,k,d,h,N)$ MSR code fail, denoted by $\mathcal{F}=\{i_0,\dots,i_{h-1}\}$.
	During the repair process of the $h$ failed nodes, $k\le d\le n-h$ helper nodes are connected and thus $n-d-h$ nodes are unconnected,
	denoted by  $\mathcal{H}=\{j_0,\dots,j_{d-1}\}\subseteq[0,n)\backslash \mathcal{F},|\mathcal{H}|=d$ and $\mathcal{U}=[0,n)\backslash(\mathcal{F}\cup \mathcal{H})=\{i_h,\dots,i_{n-d-1}\},|\mathcal{U}|=n-d-h$, respectively.
	
	In the original permutation matrix-based $(n,k,d,h=1,s^n)$  MSR code $\mathcal{C}'$, each node $i\in[0,n)$ stores a column vector  $\mathbf{f}_i=(f_{i,0},\dots,f_{i,s^n-1})^\top$ of length $s^n$, where $s=d-k+1$.
	Generating $d-k+h$ instances of $\mathcal{C}'$ by space sharing technique, we  obtain an $(n,k,d,h,N)$  MSR code with sub-packetization  $N=(d-k+h)s^n$.
	\begin{Construction}\label{con}
		For given integers $n>d \ge k>0$, $h\geq 2$, $r=n-k$, $s=d-k+1$, and $N=(d-k+h)s^{n}$,
		let $\mathbb{F}_q,q\ge n+1$ be a finite field.
		Consider an $(n,k,d,h,N)$ MSR code $\mathcal{C}$ defined by the following parity check equation
		\begin{eqnarray*}\label{Matrix form_ss}
			\underbrace{\left(\begin{array}{ccccc}
					A'_{0,0} & A'_{0,1} & \cdots  &   A'_{0,n-1}      \\
					A'_{1,0} & A'_{1,1} & \cdots  &   A'_{1,n-1}      \\
					\vdots  & \vdots  &  \ddots  &  \vdots          \\
					A'_{r-1,0} & A'_{r-1,1} & \cdots  &   A'_{r-1,n-1}      \\
				\end{array}\right)}_{\mathrm{block ~matrix~}  }
			\left(\begin{array}{ccccc}
				\mathbf{f}_0        \\
				\mathbf{f}_1\\
				\vdots\\
				\mathbf{f}_{n-1}\\
			\end{array}
			\right)=\mathbf{0},
		\end{eqnarray*}
		therein, for any $i\in[0,n)$ and $t\in[0,r)$,
		\begin{eqnarray}\label{matrixA'}
			A'_{t,i}\triangleq\begin{pmatrix}
				A_{t,i} &0 &0 &0 \\
				0 &A_{t,i}&0 &0 \\
				\vdots &\vdots &\ddots &\vdots \\
				0 &0 &0 &A_{t,i}
			\end{pmatrix}\in \mathbb{F}^{N\times N}_q,
			%	\end{split}
	\end{eqnarray}
	where $A_{t,i}$ is an $s^n\times s^n$ matrix defined in \eqref{A_i^t},
	$\mathbf{f}_i=((\mathbf{f}_i^{(0)})^\top, \dots,(\mathbf{f}_i^{(d-k+h-1)})^\top)^\top$ is a column vector of length $(d-k+h)s^n$ stored at node $i\in[0,n)$,
	and $\mathbf{f}_{i}^{(w)}=\{f_{i,a}^{(w)}:a\in[0,s^n)\}^\top, i\in[0,n), w\in[0,d-k+h)$ is a column vector of length $s^n$.
	
	In this approach, the codeword $\mathbf{f}_i,i\in[0,n)$ is composed of $\mathbf{f}_{i}^{(w)},w\in[0,d-k+h)$. Formally, each component  $\mathbf{f}_{i}^{(w)}$ is referred to as an instance of $\mathcal{C}$. In the literature,  this process for generating $\mathbf{f}_i$ from multiple instances is known as space sharing.

\end{Construction}
%Let $\gamma$ be the primitive element over $\mathbb{F}_q, q\ge n+1$.

%	\begin{Lemma}
	%		The code $\mathcal{C}$ given by Construction \ref{con} is an MDS array code.
	%	\end{Lemma}
%\begin{proof}
%	The MDS property of code $\mathcal{C}$ can be proved similarly to Theorem 15 in \cite{ye2017explicit}, we omit the details here.
%\end{proof}

\begin{Remark}
	Comparing the parity matrix \eqref{matrixA'} of our constructions with \eqref{A_i^t} of the original $(n,k,d,h=1,s^n)$  MSR  code $\mathcal{C}'$, it is not difficult to find our constructions consists of $d-k+h$ instances of the code $\mathcal{C}'$. We only extend the length of column vectors stored across each node, without altering the structure of the parity matrix.
	This means that the parity check matrix  \eqref{matrixA'} is still a permutation matrix.
\end{Remark}%Therefore, the codes proposed in Construction \ref{con} are indeed Zigzag MSR codes over $\mathbb{F}_q,q\ge n+1$.}

%In what follows, we introduce a cooperative repair scheme for array codes by Construction \ref{con}.
%Assume that $h$ nodes of an $(n,k,d,h,N)$ MSR code fail, denoted by $\mathcal{F}=\{i_0,\dots,i_{h-1}\}$.
%During the repair process of the $h$ failed nodes, $k\le d\le n-h$ helper nodes are connected and thus $n-d-h$ nodes are unconnected,
%denoted by  $\mathcal{H}=\{j_0,\dots,j_{d-1}\}\subseteq[0,n)\backslash \mathcal{F},|\mathcal{H}|=d$ and $\mathcal{U}=[0,n)\backslash(\mathcal{F}\cup \mathcal{H})=\{i_h,\dots,i_{n-d-1}\},|\mathcal{U}|=n-d-h$, respectively.
%The scheme is mainly two parts: the downloading phase and the cooperative phase.
%
%{\bf Downloading phase:} For each failed node $i_u \in \mathcal{F}$ with $u \in [0, h)$, it downloads auxiliary data $\mathcal{D}_{i_u, j}$ from the helper node $j \in \mathcal{H}$, where each helper node determines the auxiliary data using Algorithm~\ref{}.
%
%\begin{algorithm}[t]
%  \DontPrintSemicolon
%  \SetKwInOut{Input}{Input}
%  \Input{
%    $\mathcal{F}$, $i_u$, $j$ and ${\bm f_j}$;
%  }
%  %\State {}
%  \Return {}
%  \caption{Auxiliary data from helper nodes}\label{algorithm_find_V_1}
%\end{algorithm}

Following \eqref{Parity-check equation}, the parity check equations of an $(n,k,d,h,N)$ MSR code $\mathcal{C}$ with sub-packetization  $N=(d-k+h)s^n$ in Construction \ref{con} can be given as
\begin{align*}
A_{t,0}\mathbf{f}_0^{(w)}+A_{t,1}\mathbf{f}_1^{(w)}+\cdots+A_{t,n-1}\mathbf{f}_{n-1}^{(w)} = \mathbf{0},
\end{align*}
where  $w\in[0,d-k+h),t\in[0,r)$, $A_{t,i}$ and $A'_{t,i},i\in[0,n)$ are defined by \eqref{A_i^t} and \eqref{matrixA'}, respectively.

According to \eqref{eqn:a-row}  and \eqref{matrixA'}, for fixed $w\in[0,d-k+h)$, the $a$-th row of the $t$-th parity check equation of  $(n,k,d,h,N=(d-k+h)s^n)$ MSR code $\mathcal{C}$ is
\begin{eqnarray}\label{eqn:a-th parity}
\sum_{i=0}^{n-1}A_{t,i}(a)\mathbf{f}_i^{(w)} =
\sum_{i=0}^{n-1}\zeta_{t,i,a}f_{i,a(i,t)}^{(w)} = 0,
\end{eqnarray}
where $a\in[0,s^n),t\in[0,r),s=d-k+1$.

In what follows, we propose the cooperative repair scheme for the  $h$ failed nodes of the $(n,k,d,h,N)$ MSR code $\mathcal{C}$ with sub-packetization  $N=(d-k+h)s^n$ with $s=d-k+1$. By analyzing the parity check equations in \eqref{eqn:a-th parity}, we observe that the unknowns in the failed nodes can be recovered using a classical  interference alignment technique. To this aim, for each failed node, we first sum $d-k+1$ instances of the MSR code $\mathcal{C}$ and then group all symbols to subgroups according to some specific rules. These two steps ensure the perfect interference alignment, enabling the failed nodes to be recovered through the Download and Cooperative phases. Specifically, the repair process is composed of the following four steps.

\textbf{\textit{ Step 1 (Summing):}} For each failed node $i_u,u\in[0,h)$, respectively sum the parity check equations of $s=d-k+1$ instances to obtain $r\cdot s^n$ equations.

In order to repair the failed node $i_u\in\mathcal{F}=\{i_0,\dots,i_{h-1}\}$, $u\in[0,h)$, let $a=a(i_u,w),w\in[0,s)$ in \eqref{eqn:a-th parity}. Then, for given $t\in[0,r)$ and $a\in [0,s^n)$,
we sum
\begin{eqnarray*}\label{eqn:w parity}
&&\zeta_{t,i_u,a(i_u,w)}
f_{i_u,a(i_u,w+t)}^{(w)}\\
&+&\sum_{i\in[0,n)\backslash\{i_u\}}\zeta_{t,i,a(i_u,w)}
f_{i,a(i_u,w;i,t)}^{(w)}= 0,
w\in[0,s-2]
\end{eqnarray*}
and
\begin{eqnarray*}\label{eqn:s+u-1 parity}
&&\zeta_{t,i_u,a(i_u,s-1)}
f_{i_u,a(i_u,s-1+t)}^{(u+s-1)}\\
&+&\sum_{i\in[0,n)\backslash\{i_u\}}\zeta_{t,i,a(i_u,s-1)}
f_{i,a(i_u,s-1;i,t)}^{(u+s-1)}= 0
\end{eqnarray*}
to obtain
%\begin{footnotesize}
\begin{eqnarray}\label{Eqn: summing parity}
&&\sum_{w=0}^{s-2}	\zeta_{t,i_u,a(i_u,w)}
f_{i_u,a(i_u,w+t)}^{(w)}\nonumber\\
&+&	\zeta_{t,i_u,a(i_u,s-1)}
f_{i_u,a(i_u,s-1+t)}^{(u+s-1)}\nonumber\\
&+&
\sum_{i\in[0,n)\backslash\{i_u\}}\biggr(\sum_{w=0}^{s-2}\zeta_{t,i,a(i_u,w)} f_{i,a(i_u,w;i,t)}^{(w)}\nonumber\\
&+&\zeta_{t,i,a(i_u,s-1)}
f_{i,a(i_u,s-1;i,t)}^{(u+s-1)}\biggr)=0,
\end{eqnarray}
which is called  the \textit{sum parity check equation $(a,t)$  of node $i_u$}, or simply SPCE   $(a,t)$  for short. In total, there are
$r\cdot s^n$ SPCEs.
We recall that $a(i_u,w)=(a_0,\dots,\langle a_{i_u}+ w\rangle,\dots,a_{n-1})$ and $a(i_u,w;i,t)=(a_0,\dots,\langle a_{i_u}+w\rangle,\dots,\langle a_{i}+t\rangle,\dots,a_{n-1})$, where  $\langle \cdot \rangle$ means the smallest nonnegative residue modulo $s$.

For ease of analysis in what follows, we rewrite  \eqref{Eqn: summing parity} as
\begin{eqnarray}\label{eqn:simply0}
&&	\sum_{w=0}^{s-2}\zeta_{t,i_u,a(i_u,w)}
f_{i_u,a(i_u,w+t)}^{(w)}\nonumber\\
&&+
\zeta_{t,i_u,a(i_u,s-1)}
f_{i_u,a(i_u,s-1+t)}^{(u+s-1)}\nonumber\\
&&+\sum_{i\in\mathcal{F}\cup\mathcal{U}\backslash\{i_u\}}\biggr(\sum_{w=0}^{s-2}\zeta_{t,i,a(i_u,w)}f_{i,a(i_u,w;i,t)}^{(w)}\nonumber\\
&&+\zeta_{t,i,a(i_u,s-1)}f_{i,a(i_u,s-1;i,t)}^{(u+s-1)}\biggr)\nonumber\\
&=&\sum_{w=0}^{s-2}\zeta_{t,i_u,a(i_u,w)}f_{i_u,a(i_u,w+t)}^{(w)}\nonumber\\
&&+
\zeta_{t,i_u,a(i_u,s-1)}
f_{i_u,a(i_u,s-1+t)}^{(u+s-1)}\nonumber\\
&&+\sum_{i\in\mathcal{F}\cup\mathcal{U}\backslash\{i_u\}}\zeta_{t,i,a}\left(\sum_{w=0}^{s-2}f_{i,a(i_u,w;i,t)}^{(w)}+f_{i,a(i_u,s-1;i,t)}^{(u+s-1)}\right)\nonumber\\
&=&-\sum_{j\in\mathcal{H}}\zeta_{t,j,a}\left(\sum_{w=0}^{s-2}f_{j,a(i_u,w;j,t)}^{(w)}+f_{j,a(i_u,s-1;j,t)}^{(u+s-1)}\right),
\end{eqnarray}
where $a\in[0,s^n),t\in[0,r)$, and the identities follow from
\begin{eqnarray*}
\zeta_{t,i,a}= \zeta_{t,i,a(i_u,1)}=\cdots = \zeta_{t,i,a(i_u,s-1)},\quad
i\in[0,n)\backslash \{i_u\}
\end{eqnarray*}
by  P1 and P2,
indicating that although the $i_u$-th digit of $a$ changes, $\zeta_{t,i,a}$ only cares about the $i$-th digit of $a$.

\textbf{\textit{Step 2 (Grouping):}}  Divide all the sum parity check equations in \eqref{eqn:simply0} into $s^{d+1}$ groups with $s=d-k+1$, each group having $r\cdot s^{n-d-1}$ equations.

Given  $\mathcal{H}=\{j_0,\dots,j_{d-1}\}$  and $\ell=(\ell_0,\dots,\ell_{d-1})\in[0,s^d)$, where $\ell_i\in[0,s),i\in[0,d)$,
define
\begin{eqnarray}\label{eqn:a}
\mathcal{A}_{\ell}\triangleq \{a:a_{j_m}=\ell_m,j_m\in\mathcal{H},m\in[0,d)\}\subseteq [0,s^n),%\nonumber \\
%\ell \in[0,s^d),
\end{eqnarray}
and
\begin{eqnarray}\label{eqn:b}
\mathcal{B}_{\ell,\tau}&\triangleq&\{a: a_{i_0} +a_{i_1}+\cdots+a_{i_{n-d-1}}
\equiv \tau \mod s\}  \nonumber \\
&\subseteq& \mathcal{A}_{\ell}, \quad \tau\in [0,s),
\end{eqnarray}
where $i_m\in[0,n)\backslash\mathcal{H},m\in[0,n-d)$.
It is obvious that $\cup_{\ell=0}^{s^d-1}\cup_{\tau=0}^{s-1}\mathcal{B}_{\ell,\tau} =\cup_{\ell=0}^{s^d-1}\mathcal{A}_{\ell}= [0,s^n)$ and
$ \mathcal{B}_{\ell,\tau}$ has $s^{n-d-1}$ elements for any $\ell\in[0,s^d)$ and $\tau\in [0,s)$.

When forming subgroups in \eqref{eqn:a} and \eqref{eqn:b}, we classify all symbols $a\in[0,s^n)$ by its digits corresponding to helper nodes first and group $\mathcal{A}_{\ell}$ by looking at the sum of the digits corresponding to failed and unconnected nodes. This is so that $a(i,t)$ all lie in the same subgroup as long as $i\in\mathcal{I}$.

Then,  we partition all the $r\cdot s^n$ SPCEs in \eqref{eqn:simply0} into $s^{d+1}$ groups such that
each group includes  $s^{n-d-1}$ subgroups with each subgroup containing $r$ SPCEs. Precisely, given $\ell\in [0,s^{d})$  and $\tau\in [0,s)$, the $(\ell,\tau)$-th group
consists of SPCEs  $\{(a(i_u,-t),t):t\in [0,r)\}$ for all $\mathbf{a}\in \mathcal{B}_{\ell,\tau}$, i.e.,
\begin{eqnarray}\label{eqn:simply1}
&&	\sum_{w=0}^{s-2}\zeta_{t,i_u,a(i_u,w-t)}
f_{i_u,a(i_u,w)}^{(w)}\nonumber \\
&&+
\zeta_{t,i_u,a(i_u,s-1-t)}
f_{i_u,a(i_u,s-1)}^{(u+s-1)}\nonumber\\
&&+\sum_{i\in\mathcal{F}\cup\mathcal{U}\backslash\{i_u\}}\zeta_{t,i,a(i_u,-t)}\biggr(\sum_{w=0}^{s-2}f_{i,a(i_u,w-t;i,t)}^{(w)}\nonumber \\
&&+f_{i,a(i_u,s-1-t;i,t)}^{(u+s-1)}\biggr)\nonumber\\
\nonumber\\
&=&-\sum_{j\in\mathcal{H}}\zeta_{t,j,a(i_u,-t)}\biggr(\sum_{w=0}^{s-2}f_{j,a(i_u,w-t;j,t)}^{(w)}\nonumber \\
&&+f_{j,a(i_u,s-1-t;j,t)}^{(u+s-1)}\biggr),
\end{eqnarray}
where given $\mathbf{a}\in \mathcal{B}_{\ell,\tau}$, the $r$ SPCEs form a \textit{subgroup}.

As shown in \eqref{eqn:simply1}, the data on the left-hand side (LHS) of \eqref{eqn:simply1} corresponds to $h$ failed nodes and $n-d-h$ unconnected nodes. According to the grouping rule specified in \eqref{eqn:b}, for a fixed $a\in\mathcal{B}_{\ell,\tau}$, where $\ell \in [0, s^d),\tau \in [0, s)$, the data
$$\left\{f_{i_u,a(i_u,m)}^{(w)}:a \in \mathcal{B}_{\ell,\tau},m \in [0, s),w\in[0,d-k+h)\right\}$$ and
$$\left(\sum_{w=0}^{s-2}f_{i,a(i_u,w-t;i,t)}^{(w)}+f_{i,a(i_u,s-1-t;i,t)}^{(u+s-1)}\right)$$
remain within the same subset $\mathcal{B}_{\ell,\tau}$ as $t$ ranges over $[0, r)$.
This implies that
\begin{itemize}
\item[G.] given $\ell \in [0, s^d)$ and $\tau \in [0, s)$, the unknowns within a subgroup are confined to the $(\ell,\tau)$-th group.\label{E1}
\end{itemize}

G holds because the set $\mathcal{A}_{\ell}$ depends on the digits $a_j, j \in \mathcal{H}$, whereas the unknowns in \eqref{eqn:simply1} correspond to $a(i_u, \star)$ with $i_u \notin \mathcal{H}$. Similarly, the set $\mathcal{B}_{\ell,\tau}$ is determined by the sum of certain digits, which remains unchanged since the same value, $t$, is added to one of the digits involved in the summation.

Consequently, by fixing  $\ell\in[0,s^d)$ and $\tau\in[0,s)$,
we achieve perfect interference alignment within each set $\mathcal{B}_{\ell,\tau}$ via \eqref{eqn:simply1}, which guarantees that the desired data can be recovered by downloading the data  on the right hand side (RHS) of \eqref{eqn:simply1} as follows.

\textbf{\textit{Step 3 (Download phase):}} By downloading
\begin{eqnarray}\label{Eqn_Dd_data}
\mathcal{D}_j=f_{j,a(i_u,0)}^{(0)}+\cdots+f_{j,a(i_u,s-2)}^{(s-2)}+f_{j,a(i_u,s-1)}^{(u+s-1)},\nonumber \\
a\in [0,s^n)
\end{eqnarray}
from all the helper nodes $j\in\mathcal{H}$, where $s=d-k+1$,   the failed node $i_u,u\in[0,h)$ solves SPCEs in
the $(\ell, \tau)$-th group to
respectively recover
\begin{eqnarray}\label{Eqn_Re_data_1}
\left\{f_{i_u,a}^{(0)},\dots,f_{i_u,a}^{(s-2)},f_{i_u,a}^{(u+s-1)}:a\in[0,s^n)\right\}
\end{eqnarray}
and
\begin{align}\label{Eqn_Re_data_2}
\Big\{f_{i_v,a(i_u,0)}^{(0)}+\cdots+f_{i_v,a(i_u,s-2)}^{(s-2)}+f_{i_v,a(i_u,s-1)}^{(u+s-1)}:\nonumber \\
a\in [0,s^n),\forall~ i_{v}\in \mathcal{F}\cup \mathcal{U}\backslash\{i_u\}\Big\},
\end{align}
where $a(i_u,0)$ represents $a\in[0,s^n)$. For notation consistency throughout our formulation, we employ $a(i_u,0)$ hereafter.

\textbf{\textit{Step 4 (Cooperative phase):}} The failed node $i_u$ recovers the remaining data after receiving
\begin{eqnarray}\label{Eqn_Co_data}
f_{i_u,a(i_v,0)}^{(0)}+\cdots+f_{i_u,a(i_v,s-2)}^{(s-2)}+f_{i_u,a(i_v,s-1)}^{(v+s-1)}
, a\in [0,s^n)
\end{eqnarray}
from the failed nodes
$i_{v}, v\in [0,h)\backslash\{u\}$.

As demonstrated in Step 3, all failed nodes will recover the first $s-1$ instances, with one instance in the last $h$ instances of itself during the Download phase, and a linear combination of different instances of other failed nodes.
Then, in Step 4, they help each other reconstruct the last $h$ instances during the Cooperative phase. Note that, naturally, different instances do not interact with each other, then all failed data is recovered.

%Before proving that Steps 3 and 4 are feasible, we  first provide an  example to demonstrate the repair process of $h=2$ failed nodes as follows.

After Summing and Grouping steps,  note from Fact G that for  fixed $\ell \in [0, s^d)$ and $\tau \in [0, s)$, there are  $r\cdot s^{n-d-1}$ unknowns across the $r\cdot s^{n-d-1}$ SPCEs \eqref{eqn:simply1} in the $(\ell,\tau)$-th group of node $i_u$.
In what follows, we show these two steps  precisely achieve the interference alignment by  Lemma \ref{lemma:recover}.
To achieve this, we first set a  logical order to the $h$ failed nodes and $n-d-h$ unconnected nodes, still denote $\mathcal{F}\cup \mathcal{U}=\{i_0,i_1,\dots,i_{n-d-1}\}$ with $i_0<i_1<\cdots <i_{n-d-1}$.
The key insight lies in reformulating  \eqref{eqn:simply1} into a matrix form, as detailed in the following lemma. Before proceeding, for given $\ell \in [0,s^{d})$ and $\tau\in [0,s=d-k+1)$,
we define  $\mathcal{B}_{\ell,\tau}\triangleq\{\delta_{m}^{(\ell,\tau)}:m\in[0,s^{n-d-1})\}$.

\begin{Lemma}\label{lemma:recover}
For given $\ell \in [0,s^{d})$ and $\tau\in [0,s=d-k+1)$,  when fixing the failed node $i_u\in\mathcal{F}, u\in[0,h)$, the  $(r s^{n-d-1})\times (rs^{n-d-1})$ coefficient matrix of LHS of \eqref{eqn:simply1} is the
\textbf{\textit{recovery matrix}}
\begin{equation}\label{eqn:repair matrix0}
	\setlength{\arraycolsep}{1.2pt}
	\resizebox{.95\hsize}{!}{$
		R =\left(\begin{array}{ccccccccccccccc}
			%	\mathcal{F}{q,0} & \mathcal{F}{q,1} &  \cdots & \mathcal{F}{q,s-1} & \mathcal{F}_{i_1}(q) \\
			%	\hline\\
			D  & \mathbf{0} & \cdots &\mathbf{0} & B_{0,i_0} & \cdots & B_{0,i_{u-1}}& B_{0,i_{u+1}} & \cdots   & B_{0,i_{n-d-1}}\\
			\mathbf{0} & D&  \cdots &\mathbf{0} & B_{1,i_0}  & \cdots & B_{1,i_{u-1}}& B_{1,i_{u+1}} & \cdots   & B_{1,i_{n-d-1}}\\
			\vdots &  \vdots & \ddots & \vdots & \vdots & \ddots & \vdots & \vdots  & \ddots &\vdots \\
			\mathbf{0}  & \mathbf{0} & \cdots &D   &B_{s^{n-d-1}-1,i_0} &\cdots & B_{s^{n-d-1}-1,i_{u-1}}& B_{s^{n-d-1}-1,i_{u+1}} & \cdots & B_{s^{n-d-1}-1,i_{n-d-1}}\\
		\end{array}\right),
		$}\end{equation}
%\begin{figure*}
%	\begin{IEEEeqnarray}{c}\label{eqn:repair matrix0}
	%	%	\setlength{\arraycolsep}{0.3pt}
	%		R =\left(\begin{array}{ccccccccccccccc}
		%			%	\mathcal{F}{q,0} & \mathcal{F}{q,1} &  \cdots & \mathcal{F}{q,s-1} & \mathcal{F}_{i_1}(q) \\
		%			%	\hline\\
		%			D  & \mathbf{0} & \cdots &\mathbf{0} & B_{0,i_0} & \cdots & B_{0,i_{u-1}}& B_{0,i_{u+1}} & \cdots   & B_{0,i_{n-d-1}}\\
		%			\mathbf{0} & D&  \cdots &\mathbf{0} & B_{1,i_0}  & \cdots & B_{1,i_{u-1}}& B_{1,i_{u+1}} & \cdots   & B_{1,i_{n-d-1}}\\
		%			\vdots &  \vdots & \ddots & \vdots & \vdots & \ddots & \vdots & \vdots  & \ddots &\vdots \\
		%			\mathbf{0}  & \mathbf{0} & \cdots &D   &B_{s^{n-d-1}-1,i_0} &\cdots & B_{s^{n-d-1}-1,i_{u-1}}& B_{s^{n-d-1}-1,i_{u+1}} & \cdots & B_{s^{n-d-1}-1,i_{n-d-1}}\\
		%		\end{array}\right).
	%	\end{IEEEeqnarray}
%\end{figure*}
where $D$  is a $r\times s$ matrix given by
\begin{equation}\label{eqn:D}\setlength{\arraycolsep}{1.2pt}
	\resizebox{.95\hsize}{!}{$
		D=	\left(\begin{array}{cccccc}
			1 & 1 & \cdots & 1 & 1  \\
			\lambda_{i_u,1} & \lambda_{i_u,2} & \cdots &\lambda_{i_u,s-1} &\lambda_{i_u,0}  \\
			\vdots& \vdots & \ddots & \vdots& \vdots \\
			\prod\limits_{j\in[0,r-2]}\lambda_{i_u,\langle j-r+3\rangle} &
			\prod\limits_{j\in[0,r-2]}\lambda_{i_u,\langle j-r+4\rangle} & \cdots &\prod\limits_{j\in[0,r-2]}\lambda_{i_u,\langle j-r +s+1\rangle}& \prod\limits_{j\in[0,r-2]}\lambda_{i_u,\langle j-r +2\rangle}  \\
		\end{array}\right)
		$}\end{equation}
%
%		\begin{figure*}
	%			\begin{IEEEeqnarray}{c}\label{eqn:D}
		%		D=	\left(\begin{array}{cccccc}
			%			1 & 1 & \cdots & 1 & 1  \\
			%			\lambda_{i_u,1} & \lambda_{i_u,2} & \cdots &\lambda_{i_u,s-1} &\lambda_{i_u,0}  \\
			%			\vdots& \vdots & \ddots & \vdots& \vdots \\
			%			\prod\limits_{j\in[0,r-2]}\lambda_{i_u,\langle j-r+3\rangle} &
			%			\prod\limits_{j\in[0,r-2]}\lambda_{i_u,\langle j-r+4\rangle} & \cdots &\prod\limits_{j\in[0,r-2]}\lambda_{i_u,\langle j-r +s+1\rangle}& \prod\limits_{j\in[0,r-2]}\lambda_{i_u,\langle j-r +2\rangle}  \\
			%		\end{array}\right).
		%	\end{IEEEeqnarray}
	%\end{figure*}
	and
	\begin{eqnarray*}
		B_{m,i_{v}} =\big( B_{m,i_{v}}(t,j)\big)_{r\times s^{{n-d-1}}} ,\quad i_v\in\mathcal{F}\cup\mathcal{U}\backslash\{i_u\},\\m\in[0,s^{n-d-1}),j\in[0,s^{n-d-1}),\,t\in[0,r)
	\end{eqnarray*}
	is a $r\times s^{n-d-1}$ matrix with each row having only one nonzero element, i.e.,
	\begin{eqnarray}\label{eqn:BIJ}
		B_{m,i_{v}}(t,j)=\left\{\begin{array}{lll}
			\zeta_{t,i_{v},\delta_{m}^{(\ell,\tau)}(i_u,-t)},  \\
			\hspace{0.6cm}\mathrm{~~if~} j=m(v,t)\mathrm{~~and~} v<u,\\
			\zeta_{t,i_{v},\delta_{m}^{(\ell,\tau)}(i_u,-t)},  \\
			\hspace{0.6cm}\mathrm{~~if~} j=m(v-1,t)\mathrm{~~and~} v >u,\\
			0, \hspace{0.3cm}\mathrm{~~else}
		\end{array}
		\right.
	\end{eqnarray}
	with $\mathbf{m}=(m_0,m_1,\dots,m_{n-d-2})\in[0,s^{n-d-1})$, $j\in[0,s^{n-d-1})$, $t\in[0,r)$, and $\delta_{m}^{(\ell,\tau)}\in\mathcal{B}_{\ell,\tau}$ enumerates all symbols in the set $\mathcal{B}_{\ell,\tau}$ with $m$ ranging over $[0,s^{n-d-1})$.
\end{Lemma}

\begin{proof}
	The proof is given in Appendix.
\end{proof}

According to  Lemma \ref{lemma:recover},   if the matrix $R$ is nonsingular,  Steps 3 and 4 are feasible so that  the failed node $i_u$ can recover all the desired data
with $\ell$ and $\tau$ enumerating  $[0,s^{d})$ and $[0,s)$.

\begin{Lemma}\label{lemma:download}
	If the recovery matrix $R$ in \eqref{eqn:repair matrix0} is nonsingular, during Step 3 the failed node $i_u,u\in[0,h)$ can recover
	the data in \eqref{Eqn_Re_data_1} and \eqref{Eqn_Re_data_2} with the help of the  downloaded symbols in \eqref{Eqn_Dd_data}.
\end{Lemma}
\begin{proof}
	Given  $i\in [0,n)$, it is easy to check $\{a(i_u,-t;i,t):a\in[0,s^n)\}=\{a:a\in[0,s^n)\}$ for each $t\in[0,r)$, which means for any $i\in[0,n)$, we have
	\begin{eqnarray}\label{eqn:down}
		&&	\left\{\sum_{w=0}^{s-2}f_{i,a(i_u,w-t;i,t)}^{(w)}+f_{i,a(i_u,s-1-t;i,t)}^{(u+s-1)}: a\in [0,s^n)\right\}\nonumber\\
		&=&\left\{\sum_{w=0}^{s-2}f_{i,a(i_u,w)}^{(w)}+f_{i,a(i_u,s-1)}^{(u+s-1)}: a\in [0,s^n)\right\}.
	\end{eqnarray}
	Equation \eqref{eqn:down} follows since both $a(i_u,w-t)$  and $a(i,t)$ act as  permutations on the digits of $a$. Consequently, they generate the same set of variables.
	Then, for any $u\in[0,h)$, after the failed node $i_u$  downloads the data in \eqref{Eqn_Dd_data} from $d$ helper nodes $j\in\mathcal{H}$, the RHS of \eqref{eqn:simply1} is known.
	
	Thus, for fixed $\ell\in [0,s^d)$ and $\tau\in [0,s)$, there are $(s+n-d-1)s^{n-d-1}=r\cdot s^{n-d-1}$ variables in the $r\cdot s^{n-d-1}$ equations
	in \eqref{eqn:simply1}, i.e.,
	\begin{eqnarray}\label{eqn:i_u}
		\left\{f_{i_u,a(i_u,0)}^{(0)},\dots,f_{i_u,a(i_u,s-2)}^{(s-2)},f_{i_u,a(i_u,s-1)}^{(u+s-1)}:a\in \mathcal{B}_{\ell,\tau}\right\}
	\end{eqnarray}
	and
	\begin{align}\label{eqn:i_v}
		\Biggl\{ \sum_{w=0}^{s-2} f_{i_v,a(i_u,w-t;i_v,t)}^{(w)} + f_{i_v,a(i_u,s-1-t;i_v,t)}^{(u+s-1)} :
		\nonumber \\
		a \in \mathcal{B}_{\ell,\tau}, \, t \in [0,r), \, \forall i_v \in \mathcal{F} \cup \mathcal{U} \setminus \{i_u\} \Biggr\},
	\end{align}
	where $s=d-k+1$ and $r=n-k$.
	Therefore, if $R$ is nonsingular,
	the failed node $i_u\in\mathcal{F}$ is able to recover all the data in \eqref{eqn:i_u} and \eqref{eqn:i_v}.
	When $\ell$ and $\tau$ enumerate  $[0,s^{d})$ and $[0,s)$, respectively, all the desirable data in \eqref{Eqn_Re_data_1} can be repaired according to $\{f_{i_u,a(i_u,m)}^{(w)}:a\in[0,s^n)\}=\{f_{i_u,a}^{(w)}:a\in[0,s^n)\}$ for any $m\in[0,s)$ and $w\in[0,d-k+h)$, while the data in \eqref{Eqn_Re_data_2} can be recovered following \eqref{eqn:down}.	
	This completes the proof.
\end{proof}

\begin{Lemma}\label{thm:coop}
	The $h$ failed nodes of an $(n,k,d,h,N)$  MSR code  $\mathcal{C}$ proposed in Construction \ref{con} can be recovered  in Step 4 with the help of the exchanging data in \eqref{Eqn_Co_data}.
\end{Lemma}
\begin{proof}
	According to Lemma \ref{lemma:download}, if $R$ is nonsingular,  the failed node $i_u\in \mathcal{F}$  has the data in \eqref{Eqn_Re_data_1} and the failed node $i_v,v\in[0,h)\backslash\{u\}$ possesses the data in \eqref{Eqn_Co_data} after Step 3. Following that, the node $i_u$ can recover $$\left\{f_{i_{u,a}}^{(v+s-1)}:a\in[0,s^n)\right\}=\left\{f_{i_{u,a(i_v,s-1)}}^{(v+s-1)}:a\in[0,s^n)\right\}$$
	from \eqref{Eqn_Co_data} by means of the data $$\left\{f_{i_u,a(i_u,0)}^{(0)},\dots,f_{i_u,a(i_u,s-2)}^{(s-2)}:a\in[0,s^n)\right\}$$
	recovered in \eqref{Eqn_Re_data_1}.
	When $v$ enumerates $[0,h)\backslash\{u\}$,  node $i_u$ obtains all the data
	\begin{eqnarray*}
		&&\left\{f_{i_u,a}^{(w)}:w\in[0,s-2],a\in[0,s^n)\right\}\cup\Big\{f_{i_u,a}^{(u+s-1)}:\\
		&&a\in[0,s^n)\Big\}\cup\left\{f_{i_u,a}^{(v+s-1)}:v\in[0,h)\backslash\{u\},a\in[0,s^n)\right\}\\
		&=&\left\{f_{i_u,a}^{(w)}:w\in[0,h+s-2],a\in[0,s^n)\right\}\\
		&=&\left\{f_{i_u,a}^{(w)}:w\in[0,d-k+h),a\in[0,s^n)\right\}.
	\end{eqnarray*}
	That is, each failed node $i_u\in \mathcal{F}$ can be recovered.
\end{proof}

Based on Lemmas \ref{lemma:download} and \ref{thm:coop}, the following main result is evident.

\begin{Theorem}\label{P-invert}
	If all the recovery matrices $R$ in \eqref{eqn:repair matrix0} are nonsingular for any $i_u\in \mathcal{F}$, then
	the corresponding $h$ failed nodes of the $(n,k,d,h,N)$ MSR code  $\mathcal{C}$  with sub-packetization $N=(d-k+h)s^n$ generated in Construction \ref{con}  can be optimally cooperative repaired by the four Steps above, where $s=d-k+1$ is an integer.
\end{Theorem}
\begin{proof}
	By Lemmas \ref{lemma:download} and \ref{thm:coop}, the failed nodes can be recovered by designed scheme when matrices $R$ in \eqref{eqn:repair matrix0}
	are nonsingular for each $i_u\in \mathcal{F}$. We now proceed to determine the repair bandwidth.
	During the download phase, according to Lemma \ref{lemma:download},
	each failed node downloads $s^n=N/(d-k+h)$ symbols from each of the $d$ helper nodes,
	i.e., the repair bandwidth in this phase is
	\begin{eqnarray*}
		\gamma_1= hd\cdot{N\over d-k+h}.
	\end{eqnarray*}
	
	Next, during the cooperative phase,
	according to Lemma \ref{thm:coop},
	each failed node downloads $s^n=N/(d-k+h)$ symbols from the other failed nodes,
	i.e., the repair bandwidth in this phase is
	\begin{eqnarray*}
		\gamma_2 = h(h-1)\cdot{N\over d-k+h}.
	\end{eqnarray*}
	
	Totally, the repair bandwidth is
	\begin{equation*}
		\begin{split}
			\gamma =\gamma_1+\gamma_2=\frac{h(d+h-1)N}{d-k+h},
		\end{split}
	\end{equation*}
	attaining the optimal repair bandwidth given in \eqref{eqn:Lower_Bound}, which finishes the proof.
\end{proof}

\begin{Example}\label{Example_2}
	%Continue with Example \ref{Example_1}.
	Let $h=2$ and generate $d-k+h=3$ instances of the $(n=6,k=2,d=3,h=1,N=2^6)$ MSR code $\mathcal{C}'$ over $\mathbb{F}_7$.
	In this way, we  obtain an $(n=6,k=2,d=3,h=2,N=3\cdot2^6)$ MSR code $\mathcal{C}$.  Denote  the column vector of length $N=3\cdot2^6$ stored at node $i$ by $\mathbf{f}_i=((\mathbf{f}_i^{(0)})^\top,(\mathbf{f}_i^{(1)})^\top, (\mathbf{f}_i^{(2)})^\top)^\top$, where $\mathbf{f}_{i}^{(w)}=\{f_{i,a}^{(w)}:a\in[0,63]\}^\top, i\in[0,5], w\in[0,2]$ is a column vector of length $64$.
	
	Assume that the nodes $0$ and $1$ fail, and the helper nodes are $j\in[3,5]$, thus, the unconnected node is $2$.
	Take  the repair of the node $0$ as an example.
	
	\textbf{Step 1 (Summing):}	
	For $a\in [0,63],t\in[0,3]$, by \eqref{eqn:simply0}, we get $64$ SPCEs of node $0$ as
	\begin{align*}
		&\zeta_{t,0,a}f_{0,a(0,t)}^{(0)}+\zeta_{t,0,a(0,1)}f_{0,a(0,1+t)}^{(1)}\\
		&+\sum_{i=1}^2\zeta_{t,i,a}\left(f_{i,a(i,t)}^{(0)}+f_{i,a(0,1;i,t)}^{(1)}
		\right) \\
		=&	-\sum_{j=3}^5\zeta_{t,j,a}\left(f_{j,a(j,t)}^{(0)}+f_{j,a(0,1;j,t)}^{(1)}
		\right).
	\end{align*}

	%following \eqref{eqn:zeta}, \eqref{eqn:A_{t,i,a}} and \eqref{eqn:a-th parity}.	

	\begin{figure*}[ht]
		\centering
		\tikzsetnextfilename{2-butterflies}
		\scalefont{0.5}
		\begin{tikzpicture}[scale =1]
			% one butterfly
			\node[mybox] (nodein1+) { node 0} ;
			%\node[circle split, draw, above right=160pt and 100pt of S] (nodeout0+) { $s_{j_0}$ \nodepart{lower}  $ X^{(1)}_{out}$} ;
			\node[mybox, above =20pt  of nodein1+] (nodein2+) { node 1} ;
			\node[mybox, above =20pt of nodein2+] (nodein3+) { node 2} ;
			\node[mybox, above =20pt of nodein3+] (nodein4+) { node 3} ;
			\node[mybox, above =20pt of nodein4+] (nodein5+) { node 4} ;
			\node[mybox, above =20pt of nodein5+] (nodein6+) { node 5} ;
			
			%	\node[mybox, right =80 pt of nodein3+] (node1+) {$\ \ $} ;
			
			%	\node[mybox, right =80pt of nodein5+] (node3+) {$\ \ $} ;
			\node[mybox_vec, right =80 pt of nodein3+] (node1) {$\ \ $} ;
			\node[mybox_vec, right =80pt of nodein5+] (node3) {$\ \ $} ;
			\node[mybox_vec, right =90 pt of node1] (node1new) {} ;
			\node[mybox_vec, right =90pt of node3] (node3new) {} ;

			\path[black,->,very thick,auto]

			(nodein4+.east) edge [bend right] node {} (node3.west)
			(nodein5+.east) edge  node {} (node3.west)
			(nodein6+.east) edge [bend left] node {} (node3.west)
			(2.4,3.5) edge node {} (node1.west)
			(2.4,3) edge node {} (node1.west)
			(2.4,2.5) edge node {} (node1.west);
			%\draw[pattern= dots] (6.75,5.3) rectangle (8.6,5.74);
			%\draw[pattern= dots] (6.75,2.45) rectangle (8.6,2.86);
			\draw[dashed, ->] (5.1,3.0) to [bend right] node {} (5.1,5.5);
			\draw[dashed, ->] (5.1,6.0) to [bend left] node {} (5.1,2.6);
			\draw [|-|, very thick] (2.8,2.5) arc  (240:120:15pt);
			\draw [|-|,  thick] (3.18,2.86) -- (5.08,2.86);
			\draw [|-|,  thick] (3.18,5.7) -- (5.08,5.7);
			%\draw[pattern= dots] (11.81,5.3) rectangle (13.67,6.148);
			%\draw[pattern= dots] (11.81,2.44) rectangle (13.67,3.3);
			
			\draw node[] at (1.5,7.4) {$\color{blue}f_{3,a}^{(0)}+f_{3,a(1,1)}^{(2)}$};
			\draw node[] at (1.5,6.0) {${\color{blue}f_{4,a}^{(0)}+f_{4,a(1,1)}^{(2)}}$};
			\draw node[] at (1.5,4.0) {$\color{blue}f_{5,a}^{(0)}+f_{5,a(1,1)}^{(2)}$};
			
			\draw node[] at (4.1,5.9) {$\color{blue}f_{0,a}^{(0)}+f_{0,a(1,1)}^{(2)}$};
			\draw node[] at (4.1,5.48) {$\color{magenta}f_{1,a}^{(0)},f_{1,a}^{(2)}$};
			
			\draw node[] at (4.1,3.05) {$\color{blue}f_{1,a}^{(0)}+f_{1,a(0,1)}^{(1)}$};
			\draw node[] at (4.1,2.62) {${\color{magenta}f_{0,a}^{(0)},f_{0,a}^{(1)}}$};

			\draw node[] at (9.2,2.62) {$\color{magenta}f_{0,a}^{(0)},f_{0,a}^{(1)}$};
			\draw node[] at (9.2,3.05) {${\color{green}f_{0,a}^{(2)}}$};
			\draw node[] at (9.2,5.48) {${\color{magenta}f_{1,a}^{(0)},f_{1,a}^{(2)}}$};
			\draw node[] at (9.2,5.9) {${\color{green}f_{1,a}^{(1)}}$};
			
			\draw node[] at (2.8,2.3) {$\color{blue}f_{i,a}^{(0)}+f_{i,a(0,1)}^{(1)}$};
			
			\draw node[] at (2.2,1.5) {Download Phase};

			%	\draw [->,thick,dotted](4.4,4.4) -- (5.8,4.4);
			\draw [->,thick,dotted](6.2,4.4) -- (7.6,4.4);
			%	\draw node[] at (5,4.1) {recovering};
			\draw node[] at (6.8,4.1) {recovering};
			
			\draw node[] at (5.4,1.5) {Cooperative Phase};
			\draw [-,thick](0.25,0.25) -- (-0.25,-0.25);
			\draw [-,thick](-0.25,0.25) -- (0.25,-0.25);
			\draw [-,thick](0.25,1.65) -- (-0.25,1.15);
			\draw [-,thick](-0.25,1.65) -- (0.25,1.15);
			%\draw [-,thick](0.25,3.1) -- (-0.25,2.6);
			%\draw [-,thick](-0.25,3.1) -- (0.25,2.6);
		\end{tikzpicture}
		
		\caption{Repairing processes of failed nodes $0$ and $1$, where $i\in[3,5],a\in[0,63]$. In the download phase, failed node $u,u\in[0,1]$ downloads $f_{i,a}^{(0)}+f_{i,a(u,1)}^{(u+1)}$ from the helper node $i,i\in[3,5]$ to recover the magenta and blue data. In the cooperative phase, the blue data is exchanged in the arrow direction to repair the green data.}
		\label{fig:repairing}
	\end{figure*}
	
	\textbf{Step 2 (Grouping):}
	According to \eqref{eqn:a}, divide all the symbols $\{a:a\in[0,63]\}$ into $8$ groups, i.e.,
	\begin{eqnarray*}
		&&\mathcal{A}_0=\{0,1,2,3,4,5,6,7\},\\
		&&\mathcal{A}_1=\{8,9,10,11,12,13,14,15\},\\
		&&\mathcal{A}_2=\{16,17,18,19,20,21,22,23\},\\
		&&\mathcal{A}_3=\{24,25,26,27,28,29,30,31\},\\
		&&\mathcal{A}_4=\{32,33,34,35,36,37,38,39\},\\
		&&\mathcal{A}_5=\{40,41,42,43,44,45,46,47\},\\
		&&\mathcal{A}_6=\{48,49,50,51,52,53,54,55\},\\
		&&\mathcal{A}_7=\{56,57,58,59,60,61,62,63\}.
	\end{eqnarray*}
	Following \eqref{eqn:b}, partition the $8$ symbols in $\mathcal{A}_0$ to $2$ groups as
	$
	\mathcal{B}_{0,0}=\{0,3,5,6\},\mathcal{B}_{0,1}=\{1,2,4,7\}.
	$
	
	Take  the $(\ell=0,\tau=0)$-th group as an example.  According to \eqref{eqn:simply1}, SPCEs $\{(a(0,-t),t):a\in\mathcal{B}_{0,0},t\in[0,3]\}=\{(0,0),(3,0),(5,0),(6,0),(1,1),(2,1),(4,1),(7,1),(0,2),$ $(3,2),(5,2),(6,2),(1,3),(2,3),(4,3),(7,3)\}$ in the $(0,0)$-th group are
	\begin{align}\label{eqn:10}
		&\zeta_{t,0,a(0,-t)}f_{0,a}^{(0)}+\zeta_{t,0,a(0,1-t)}f_{0,a(0,1)}^{(1)}\nonumber\\
		&+\sum_{i=1}^2\zeta_{t,i,a(0,-t)}\left(f_{i,a(0,-t;i,t)}^{(0)}+f_{i,a(0,1-t;i,t)}^{(1)}
		\right)\nonumber\\
		=&-\sum_{j=3}^5\zeta_{t,j,a(0,-t)}\left(f_{j,a(0,-t;j,t)}^{(0)}+f_{j,a(0,1-t;j,t)}^{(1)}
		\right),
	\end{align}	
	where $ a\in \mathcal{B}_{0,0},t\in[0,3]$.

	When $t$ ranging over $[0,3]$, the identity above can be rewritten as
	$RX=MU$, where
	\[\setlength{\arraycolsep}{1.2pt}
	%  \resizebox{.95\hsize}{!}{$
		R=\left(\begin{array}{cccccccc|ccccccccc}
			1 & 1 & & & & &  & & 1 & 0 & 0 & 0& 1 & 0 & 0 & 0\\
			1 & \gamma  & & & & &  & & 0& \gamma^2 & 0 & 0 & 0 & 0& \gamma^3 & 0\\
			\gamma & \gamma  & & & &  &  &  & \gamma^2 & 0 & 0 & 0 & \gamma^3 & 0 & 0 & 0\\
			\gamma & \gamma^2  & & & & &   &  &0 & \gamma^4& 0 & 0 & 0 & 0& \gamma^6 & 0\\
			& & 1 & 1 & & & & & 0 &  1 & 0 & 0 &0 & 1 & 0 & 0\\
			& &  1 & \gamma && &  & &1 & 0 & 0 & 0 & 0 & 0 & 0 & \gamma^3\\
			& & \gamma & \gamma &&&& & 0 & \gamma^2& 0 & 0  & 0 & \gamma^3 & 0 & 0\\
			& & \gamma & \gamma^2 &&&& & \gamma^2& 0& 0 & 0& 0 & 0 & 0 & \gamma^6 \\
			
			&&& &1 & 1 & & & 0 & 0 & 1 & 0  & 0 & 0 & 1 & 0\\
			&&& &1 & \gamma & & & 0 & 0 & 0& \gamma^2 & 1 & 0 & 0 & 0 \\
			&&& &\gamma & \gamma & &  & 0 & 0 & \gamma^2 &  0 & 0  & 0 & \gamma^3 & 0\\
			&& & &\gamma & \gamma^2&  &  & 0& 0 & 0& \gamma^4 & \gamma^3 & 0 & 0 & 0 \\
			&&&&& & 1 & 1 & 0 & 0& 0 & 1  & 0 & 0& 0 & 1 \\
			&&&&& &  1 & \gamma & 0& 0 &1 & 0 & 0 &  1 & 0 & 0  \\
			&&&&& & \gamma & \gamma & 0 & 0& 0 & \gamma^2&  0 & 0  & 0 & \gamma^3\\
			&&&&& & \gamma & \gamma^2 & 0 & 0 & \gamma^2& 0 & 0 & \gamma^3 & 0 & 0
		\end{array}
		\right)%$}
	\]
	is a $16\times 16$ matrix, $X=(X_0,X_1,X_2)^\top$
	is a column vector of dimension $16$ with
	\begin{equation*}\label{eqn:failed}
		\begin{split}
			X_0&=(f_{0,0}^{(0)},f_{0,1}^{(1)},f_{0,2}^{(1)},f_{0,3}^{(0)},f_{0,4}^{(1)},f_{0,5}^{(0)},
			f_{0,6}^{(0)},f_{0,7}^{(1)}),\\ X_1&=(f_{1,0}^{(0)}+f_{1,1}^{(1)},f_{1,2}^{(1)}+f_{1,3}^{(0)},f_{1,4}^{(1)}+f_{1,5}^{(0)},f_{1,6}^{(0)}+f_{1,7}^{(1)}),\\ X_2&=(f_{2,0}^{(0)}+f_{2,1}^{(1)},f_{2,2}^{(1)}+f_{2,3}^{(0)},f_{2,4}^{(1)}+f_{2,5}^{(0)},f_{2,6}^{(0)}+f_{2,7}^{(1)}),
		\end{split}
	\end{equation*}
	and $U=(U_3,U_4,U_5)^\top$
	is a column vector of dimension $24$ with
	\begin{eqnarray*}\label{eqn:download}
		U_3&=&(f_{3,0}^{(0)}+f_{3,1}^{(1)},f_{3,9}^{(0)}+f_{3,8}^{(1)},f_{3,3}^{(0)}+f_{3,2}^{(1)},f_{3,10}^{(0)}+f_{3,11}^{(1)},\\
		&&f_{3,5}^{(0)}+f_{3,4}^{(1)},f_{3,12}^{(0)}+f_{3,13}^{(1)},f_{3,6}^{(0)}+f_{3,7}^{(1)},f_{3,15}^{(0)}+f_{3,14}^{(1)}),\\
		U_4&=&(f_{4,0}^{(0)}+f_{4,1}^{(1)},f_{4,17}^{(0)}+f_{4,16}^{(1)},f_{4,3}^{(0)}+f_{4,2}^{(1)},f_{4,18}^{(0)}+f_{4,19}^{(1)},\\
		&&f_{4,5}^{(0)}+f_{4,4}^{(1)},f_{4,20}^{(0)}+f_{4,21}^{(1)},f_{4,6}^{(0)}+f_{4,7}^{(1)},f_{4,23}^{(0)}+f_{4,22}^{(1)}),\\
		U_5&=&(f_{5,0}^{(0)}+f_{5,1}^{(1)},f_{5,33}^{(0)}+f_{5,32}^{(1)},f_{5,3}^{(0)}+f_{5,2}^{(1)},f_{5,34}^{(0)}+f_{5,35}^{(1)},\\
		&&f_{5,5}^{(0)}+f_{5,4}^{(1)},f_{5,36}^{(0)}+f_{5,37}^{(1)},f_{5,6}^{(0)}+f_{5,7}^{(1)},f_{5,39}^{(0)}+f_{5,38}^{(1)}),
	\end{eqnarray*}
	and the matrix $M$ is the $ 16\times 24$ coefficient matrix of $U$.

	Perform the following elementary  row transformation on $R$, i.e.,
	\begin{eqnarray*}
		\setlength{\arraycolsep}{1.2pt}
		\resizebox{.99\hsize}{!}{$
			R\rightarrow\left(\begin{array}{cccccccc|ccccccccc}
				1 & 1 &   &  &   &  &   &   & 1 &   &   &  & 1 &   &   &  \\
				1 & \gamma &  &   &  &   &   &   &  & \gamma^2 &   &   &   &  & \gamma^3 &  \\
				&   & 1 & 1 &  &   &  &   &  & 1 &   &   &  & 1 &   &  \\
				&   &  1 & \gamma &  &   &  &   &1 &   &   &   &   &   &   & \gamma^3\\
				&   &  &   &1 & 1 &   &  &   &   & 1 &    &   &   & 1 &  \\
				&   &  &   &1 & \gamma &   &  &   &   &  & \gamma^2 & 1 &   &   &   \\
				&   &  &   &  &   & 1 & 1 &   &  &   & 1  &   &  &   & 1 \\
				&   &  &   &  &   &  1 & \gamma  &  &   &1 &   &   &  1 &   &    \\
				&   &  &   &  &   &   &   & \gamma^2-\gamma &   &   &   & \gamma^3-\gamma &   &   &  \\
				&   &  &   &  &   &   &   &  & \gamma^4-\gamma^3&   &   &   &  & \gamma^6-\gamma^4 &  \\
				&   &  &   &  &   &  &   &   & \gamma^2-\gamma&   &    &   & \gamma^3 -\gamma&   &  \\
				&   &   &   &  &   &  &   & \gamma^2-\gamma&  &   &  &   &   &   & \gamma^6-\gamma^4 \\
				&   &  &   & &   &  &   &   &   & \gamma^2-\gamma &    &    &   & \gamma^3-\gamma &  \\
				&   &  &   &  &   &   &   &  &   &  & \gamma^4-\gamma^3 & \gamma^3-\gamma &   &   &   \\
				&   &  &   &  &   &   &   &   &  &   & \gamma^2-\gamma&    &    &   & \gamma^3-\gamma\\
				&   &  &   &  &   &  &   &   &   & \gamma^2-\gamma&   &   & \gamma^3-\gamma &   &
			\end{array}
			\right).$}
	\end{eqnarray*}
	Since the lower right submatrix
	\begin{eqnarray*}	
		\setlength{\arraycolsep}{1.2pt}
		\resizebox{.99\hsize}{!}{$\left(\begin{array}{cccccccc}
				\gamma^2-\gamma & 0 & 0 & 0 & \gamma^3-\gamma & 0 & 0 & 0\\
				0 & \gamma^4-\gamma^3& 0 & 0 & 0 & 0& \gamma^6-\gamma^4 & 0\\
				0 & \gamma^2-\gamma& 0 & 0  & 0 & \gamma^3 -\gamma& 0 & 0\\
				\gamma^2-\gamma& 0& 0 & 0& 0 & 0 & 0 & \gamma^6-\gamma^4 \\
				0 & 0 & \gamma^2-\gamma &  0 & 0  & 0 & \gamma^3-\gamma & 0\\
				0& 0 & 0& \gamma^4-\gamma^3 & \gamma^3-\gamma & 0 & 0 & 0 \\
				0 & 0& 0 & \gamma^2-\gamma&  0 & 0  & 0 & \gamma^3-\gamma\\
				0 & 0 & \gamma^2-\gamma& 0 & 0 & \gamma^3-\gamma & 0 & 0
			\end{array}
			\right)$}
	\end{eqnarray*}
	is equivalent with
	\begin{eqnarray*}
		\setlength{\arraycolsep}{1.2pt}
		\resizebox{.99\hsize}{!}{$
			\left(\begin{array}{cccccccccccccccc}
				\gamma^2-\gamma &   &   &   & \gamma^3-\gamma &   &   &  \\
				& \gamma^2-\gamma&   &    &   & \gamma^3 -\gamma&   &  \\
				&   & \gamma^2-\gamma &    &    &   & \gamma^3-\gamma &  \\
				&  &   & \gamma^2-\gamma&    &    &   & \gamma^3-\gamma\\	
				&  &   &  & \gamma-\gamma^3 &   &   & \gamma^6-\gamma^4 \\
				&   &   &   &   & \gamma^3-\gamma & \gamma-\gamma^3 &  \\
				&   &   &   &   &   & (\gamma^4-\gamma^3)(\gamma^2-1)&  \\
				&   &  &   &  &   &   & \gamma^6+\gamma^3-\gamma^5-\gamma^4 \\
			\end{array}
			\right),$}
	\end{eqnarray*}
	which is invertible.
	As a result, the equation \eqref{eqn:10} has a unique solution because of $\mathrm{Rank}(R)=16$.

	\textbf{Step 3 (Download phase):}
	Node $0$ can recover $X=(X_0,X_1,X_2)^\top$
	by downloading $U=(U_3,U_4,U_5)^\top$
	from the helper nodes $j\in[3,5]$ according to the equation \eqref{Eqn_Dd_data}.
	
	The repair process of the symbols in other sets, such as $\mathcal{B}_{0,1}$ and $\mathcal{A}_i,i\in[1,7]$ is similar, which means node $0$ can recover $f_{0,a}^{(0)},f_{0,a}^{(1)}$ and
	node $1$ can recover $f_{0,a}^{(0)}+f_{0,a(1,1)}^{(2)},a\in[0,63]$ in the download phase.
	
	\textbf{Step 4 (Cooperative phase):}
	During the cooperative phase, node $1$ transfers $f_{0,a}^{(0)}+f_{0,a(1,1)}^{(2)}$ to node $0$, such that node $0$ can recover $f_{0,a(1,1)}^{(2)}$, i.e., $f_{0,a}^{(2)},a\in[0,63]$.
	Thus, node $0$ is recovered.
	
	The repair process of node 1 follows a similar procedure to that of node 0 and is therefore omitted. Figure \ref{fig:repairing} provides an intuitive illustration of the repair processes for both failed nodes $0$ and $1$.

	\textbf{Comparison:}
	The performance of the proposed code $\mathcal{C}$ is summarized in Table \ref{tab:comparison}, along with a comparison to several representative MSR code constructions. Since all codes under consideration attain the optimal repair bandwidth $\gamma$, we focus on the sub-packetization level $N$ and the required finite field size $\mathbb{F}_q$.
	For the parameters $n=6$, $k=2$, $d=3$, and $h=2$, the proposed code achieves a sub-packetization level of  $N = 3 \cdot 2^6$, which is substantially smaller than the values reported in \cite{ye2018cooperative} and \cite{zhang2020explicit}.
	As the codes in the schemes of \cite{ye2020new}, \cite{liu2022new} and \cite{zhang2024construction} require a larger field size and  do not exist over $\mathbb{F}_{2^3}$, we include the nearest larger field size, $2^4$, for comparison. For detailed parameters, refer to Table~\ref{tab:comparison}.
	\begin{table}[h]
		\centering
		\begin{center}\caption{Comparisons of $(n=6,k=2,d=3,h=2,N)$ MSR codes with optimal cooperative repair property.}\label{tab:comparison}
			\renewcommand\arraystretch{1.2}
			\begin{tabular}{|c|c|c|c|}
				\hline
				$\gamma=h(d+h-1)N/(d-k+h)$&	 $N$  & $\mathbb{F}_q$  & Ref.\\
				\hline
				$8\cdot 3^{14}$	&$3^{15}$    &$\mathbb{F}_{2^4}$ & \cite{ye2018cooperative}\\	
				\hline
				$8\cdot 3^{14}$&$3^{15}$    &$\mathbb{F}_{2^3}$ & \cite{zhang2020explicit}\\
				\hline
				$2^9$	&$3\cdot 2^6$ &$\mathbb{F}_{2^4}$ & \cite{ye2020new}\\
				\hline
				$2^9$&	$3\cdot 2^6$ &$\mathbb{F}_{2^4}$ & \cite{liu2022new} \\
				\hline
				$2^6$	&$3\cdot 2^3$ &$\mathbb{F}_{2^4}$ &\cite{zhang2024construction}\\
				\hline
				$2^9$&	$3\cdot 2^6$ &$\mathbb{F}_{2^3}$ & The code $\mathcal{C}$ \\
				\hline
			\end{tabular}
		\end{center}
	\end{table}
\end{Example}

Theorem \ref{P-invert} and Lemma \ref{lemma:recover} indicate that any $h$ failed nodes of $(n,k,d,h,N)$  MSR codes proposed in Construction \ref{con} can be optimally cooperative repaired if the recovery matrix $R$ is nonsingular. In the next section, we will show that the matrix $R$ is indeed nonsingular when any $h$ nodes fail. 	

\section{The recovery matrix for any $h$ Failed Nodes}\label{sec:h=2,3}
%	\subsection{The recoverability of any $h$ failed nodes}	
In this section, we establish the invertibility of the recovery matrix $R$. We first demonstrate that a sub-matrix of the coefficient matrix corresponding to $h-1$ failed nodes coincides with the encoding matrix of a permutation matrix-based MSR code and is therefore invertible.
Based on this observation, we further prove that the recovery matrix is nonsingular through suitable row and column elementary transformations. Finally, we discuss practical implementation of the proposed schemes.

For an $(n,k,d,h,N)$ MSR code  $\mathcal{C}$, recall that the $h$ failed nodes are $\mathcal{F}=\{i_0,\dots,i_{h-1}\}$, and the unconnected nodes are $\mathcal{U}=\{i_h,\dots,i_{n-d-1}\}$. In this case, $r=n-k=s+n-d-1$, where $s=d-k+1$.
Before showing the invertibility of recovery matrix $R$ in \eqref{eqn:repair matrix0}, we analyze the matrix $B_{m,i_v},m\in\mathcal{R}=[0,s^{n-d-1}),i_v\in\mathcal{F}_u=\mathcal{F}\cup\mathcal{U}\backslash \{i_u\}$.
For convenience, denote $\mathcal{T}=[0,n-d-2], \mathcal{N}=[0,s^n), \mathcal{S}=[0,s)$ and $\mathcal{S}_r=[s,r)$   in the following.

Review that the recovery matrix $R$ given in  Lemma \ref{lemma:recover} is composed of coefficient matrix of the failed data at node $i_u$, i.e., Vandermonde matrix block $D$, and the coefficient matrix of the other failed nodes and the unconnected nodes $i_m\in\mathcal{F}\cup\mathcal{U}\backslash\{i_u\}$, i.e., $\mathcal{B}_{m,i_v},m\in[0,s^{n-d-1})$.
To demonstrate the invertibility of the recovery matrix $R$ in \eqref{eqn:repair matrix0}, we perform an elementary row transformation on $R$ by reordering the rows of each sub-block, such that the top-left block of the diagonal are the first $s=d-k+1$ rows of $D$, which are some Vandermonde matrix blocks, and the bottom-right block is a matrix formed by linear combinations of the first and the last $r-s$ rows of $\mathcal{B}_{t,i_m}$, which is invertible following Lemma \ref{Lemma:B}.

\begin{Lemma}\label{Lemma:B}
	For $m\in\mathcal{R}$, let	$$B_{m,\mathcal{F}_u}\triangleq(B_{m,i_0},\dots,B_{m,i_{u-1}},B_{m,i_{u+1}},\dots,B_{m,i_{n-d-1}}),$$ and
	denote the first $n-d-1$ rows of $B_{m,\mathcal{F}_u}$ by $B_{m,\mathcal{F}_u}(\mathcal{T})$.
	Then,
	$$B_{\mathcal{R},\mathcal{F}_u}(\mathcal{T})\triangleq\left(B^{\top}_{0,\mathcal{F}_u}(\mathcal{T}),\dots,
	B^{\top}_{s^{n-d-1}-1,\mathcal{F}_u}(\mathcal{T})\right)^\top$$
	is an $(n-d-1)s^{n-d-1}\times (n-d-1)s^{n-d-1}$ invertible matrix over $\mathbb{F}_q,q\ge n+1$, where $s=d-k+1$ and $\top$ is the transpose operator.
\end{Lemma}
\begin{proof}
	Firstly,
	%applying  \eqref{eqn:zeta} to \eqref{eqn:BIJ}, we have
	%\begin{eqnarray*}\label{eqn:betat-1}
	%	\zeta_{t,i_{v},\delta_{m}^{(\ell,\tau)}-t\mathbf{e}_{n,i_u}}
	%	&=&	\left\{\begin{array}{lll}
		%		1, & \text{ if } t= 0,\\
		%		\prod\limits_{0\leq j < t}\lambda_{i,\langle{(\delta_{m}^{(\ell,\tau)}-t\mathbf{e}_{n,i_u})_{i_v}}+j\rangle}, &\text{ else}
		%	\end{array}	
	%	\right.
	%\end{eqnarray*}
	%for $t\in\mathcal{T},i_v\in\mathcal{F}_u,m\in \mathcal{R}$.
	%\begin{eqnarray}\label{eqn:betat-1}
	%	\zeta_{t,i_{v},\delta_{m}^{(\ell,\tau)}-t\mathbf{e}_{n,i_u}}&= & \left\{\begin{array}{lll}	
		%		\gamma^{i_{v}+1}, & \mathrm{if}~ t\ne 0, s>n-d-2, (\delta_{m}^{(\ell,\tau)})_{i_v}+g=0~\mathrm{for ~a~} g\in[0,t),\\
		%		\gamma^{\lceil \frac{t}{s}\rceil(i_v+1)}, & \mathrm{if}~ t\ne 0, s<n-d-2,
		%		(\delta_{m}^{(\ell,\tau)})_{i_v}+g=0~\mathrm{for ~a~} g\le t\bmod s,\\
		%		\gamma^{\lfloor \frac{t}{s}\rfloor(i_v+1)}, & \mathrm{if}~ t\ne 0, s<n-d-2,
		%		 (\delta_{m}^{(\ell,\tau)})_{i_v}+g=0~\mathrm{for ~a~} g> t\bmod s,\\
		%		1, &\text{ else}
		%	\end{array}	
	%	\right.
	%\end{eqnarray}
	we show the matrix $B_{\mathcal{R},\mathcal{F}_u}(\mathcal{T})$  determined by the entry in \eqref{eqn:BIJ}, i.e.,
	\begin{eqnarray}\label{eqn:BIJ-1}
		B_{m,i_{v}}(t,j)=\left\{\begin{array}{lll}
			\zeta_{t,i_{v},\delta_{m}^{(\ell,\tau)}(i_u,-t)},  \\
			\hspace{0.6cm}\mathrm{~~if~} j=m(v,t)\mathrm{~~and~} v<u,\\
			\zeta_{t,i_{v},\delta_{m}^{(\ell,\tau)}(i_u,-t)},  \\
			\hspace{0.6cm}\mathrm{~~if~} j=m(v-1,t)\mathrm{~~and~} v >u,\\
			0, \hspace{0.3cm}\mathrm{~~else}
		\end{array}
		\right.
	\end{eqnarray}
	for all $m\in \mathcal{R}$, is equivalent to a submatrix of
	the parity check matrix of an $(n,k'=d+1,d'=2d-k+1,h,s^n)$ MSR code over $\mathbb{F}_q,q\ge n+1$. In this code, the number of parity check nodes is  $r=n-d-1$.

	According to the definition of $(n,k'=d+1,d'=2d-k+1,h,s^n)$ MSR codes in \eqref{zigzag:A_i:parameter}-\eqref{A_i^t}, the $r\cdot s^n \times r\cdot s^n$ parity check matrices of  $n-d-1$ nodes $i_v\in\mathcal{F}_u$, i.e.,
	\begin{eqnarray}\label{eqn:ATT}
		\resizebox{0.9\hsize}{!}{$	\begin{split}
				&	A_{\mathcal{T},\mathcal{F}_u} = \\
				&\left(\begin{array}{ccccccc}
					A_{0,i_0} & \cdots& A_{0,i_{u-1}} & A_{0,i_{u+1}}& \cdots & A_{0,i_{n-d-1}}\\
					A_{1,i_0} & \cdots& A_{1,i_{u-1}} & A_{1,i_{u+1}}& \cdots & A_{1,i_{n-d-1}}\\
					\vdots & \ddots& \vdots& \vdots & \ddots & \vdots\\
					A_{n-d-2,i_0} & \cdots& A_{n-d-2,i_{u-1}} & A_{n-d-2,i_{u+1}}& \cdots & A_{n-d-2,i_{n-d-1}}
				\end{array}
				\right)\end{split}
			$}
	\end{eqnarray}
	are characterized by
	\begin{eqnarray}\label{eqn:AB}	
		A_{t,i_v}(a,j)=\left\{\begin{array}{lll}
			\zeta_{t,i_v,a},  & \text{~~if~} j=a(i_v,t),\\
			0, & \text{~~else,}
		\end{array}
		\right.
	\end{eqnarray}
	where $t\in\mathcal{T},\mathbf{a}=(a_0,a_1,\dots,a_{n-1})\in\mathcal{N}$ and $j\in\mathcal{N}$.
	
	Let
	\begin{align*}
		\delta_{m}^{(\ell,\tau)}=\left((\delta_{m}^{(\ell,\tau)})_0,\dots,(\delta_{m}^{(\ell,\tau)})_{i_0},
		\dots,(\delta_{m}^{(\ell,\tau)})_{i_1},\dots,\right.\\
		\left.(\delta_{m}^{(\ell,\tau)})_{i_{n-d-1}},\dots,(\delta_{m}^{(\ell,\tau)})_{n-1}\right)
	\end{align*}
	with
	\begin{eqnarray}\label{eqn:beta_m}
		\resizebox{0.9\hsize}{!}{$	
			\begin{split}
				&\left((\delta_{m}^{(\ell,\tau)})_{i_0},(\delta_{m}^{(\ell,\tau)})_{i_1},\dots,(\delta_{m}^{(\ell,\tau)})_{i_{n-d-1}}\right)\\
				=&
				\left(m_0,m_1,\dots,m_{u-1},(\delta_{m}^{(\ell,\tau)})_{i_u},m_{u},m_{u+1},\dots,m_{n-d-2}\right).
			\end{split}
			$}
	\end{eqnarray}
	Define
	$\mathcal{L}\triangleq
	\left\{\delta_{m}^{(\ell,\tau)}: m\in\mathcal{R}\right\}\subseteq\mathcal{N}$ and $\overline{\mathcal{L}}=\mathcal{N}\backslash \mathcal{L}$.
	
	By means of reordering of  rows and columns, $A_{t,i_v}$ ($t\in\mathcal{T},i_v\in\mathcal{F}_u$) can be rewritten as
	\begin{eqnarray*}
		A_{t,i_v}\triangleq A_{t,i_v}(\mathcal{N},\mathcal{N})\rightarrow	
		\left(\begin{array}{ll}
			A_{t,i_v}(\mathcal{L},\mathcal{L}) &A_{t,i_v}(\mathcal{L},\overline{\mathcal{L}})\\
			A_{t,i_v}(\overline{\mathcal{L}},\mathcal{L}) & A_{t,i_v}(\overline{\mathcal{L}},\overline{\mathcal{L}})
		\end{array}
		\right).
	\end{eqnarray*}
	By  \eqref{eqn:AB} and  \eqref{eqn:beta_m}, we know
	\begin{eqnarray}\label{eqn:ATI}
		&&(A_{t,i_v}(\mathcal{L},\mathcal{L}))(m,j)\nonumber\\
		&=&\left\{\begin{array}{lll}
			\zeta_{t,i_v,\delta_{m}^{(\ell,\tau)}}, \hspace{0.1cm} \mathrm{~~if~} j=m(v,t) \mathrm{~~and~} v<u,\\
			\zeta_{t,i_v,\delta_{m}^{(\ell,\tau)}}, \hspace{0.1cm} \mathrm{~~if~} j=m(v-1,t)\mathrm{~~and~} v >u,\\
			0, \hspace{1.3cm} \mathrm{~~else,}
		\end{array}
		\right.
	\end{eqnarray}
	where $t\in\mathcal{T},i_v\in\mathcal{F}_u,\delta_{m}^{(\ell,\tau)}\in \mathcal{L},m\in\mathcal{R}, j\in\mathcal{R}$.
	
	%Applying \eqref{eqn:zeta} to \eqref{eqn:ATI}, we know
	%	\begin{eqnarray*}\label{eqn_ztv}
		%		\zeta_{t,i_{v},\delta_{m}^{(\ell,\tau)}}
		%		&=&	\left\{\begin{array}{lll}
			%			1, & \text{ if } t= 0,\\
			%			\prod\limits_{0\leq j < t}\lambda_{i,\langle {(\delta_{m}^{(\ell,\tau)})_{i_v}}+j\rangle},
			%			&\text{ else}
			%		\end{array}	
		%		\right.
		%	\end{eqnarray*}
	%for $t\in\mathcal{T},i_v\in\mathcal{F}_u,m\in \mathcal{R}$.
	%		It follows  P2 that
	%		$$\zeta_{t,i_{v},\delta_{m}^{(\ell,\tau)}-t\mathbf{e}_{n,i_u}}=\zeta_{t,i_v,\delta_{m}^{(\ell,\tau)}}.$$
	It follows  P2 that
	$$\zeta_{t,i_{v},\delta_{m}^{(\ell,\tau)}(i_u,-t)}=\zeta_{t,i_v,\delta_{m}^{(\ell,\tau)}}.$$
	Then, from \eqref{eqn:BIJ-1}  and \eqref{eqn:ATI} we obtain
	\begin{eqnarray}\label{eqn:abm}
		(A_{t,i_v}(\mathcal{L},\mathcal{L}))(m)=B_{m,i_v}(t),\quad t\in\mathcal{T},i_v\in\mathcal{F}_u,m\in\mathcal{R}.
	\end{eqnarray}
	%and then
	%\begin{eqnarray}\label{eqn:ABMV}
	%	A_{\mathcal{T},i_v}(\mathcal{L},\mathcal{L})(m)=B_{m,i_v}(\mathcal{T}).
	%\end{eqnarray}

	Next, note from the MDS property of $(n,k'=d+1,d'=2d-k+1,h,s^n)$ MSR codes that the matrix $A_{\mathcal{T},\mathcal{F}_u}$ in \eqref{eqn:ATT}
	is an $(n-d-1)s^n\times (n-d-1)s^n$ invertible matrix over $\mathbb{F}_q$. With respect to the row exchange, $A_{\mathcal{T},\mathcal{F}_u}$ is equivalent to
	\begin{eqnarray*}
		A_{\mathcal{T},\mathcal{F}_u}
		&\rightarrow&	\left(\begin{array}{cccccccccccc}
			A_{\mathcal{T},\mathcal{F}_u}(\mathcal{L},\mathcal{L}) & A_{\mathcal{T},\mathcal{F}_u}(\mathcal{L},\overline{\mathcal{L}}) \\
			A_{\mathcal{T},\mathcal{F}_u}(\overline{\mathcal{L}},\mathcal{L}) & A_{\mathcal{T},\mathcal{F}_u}(\overline{\mathcal{L}},\overline{\mathcal{L}})
		\end{array}	\right)\\
		&=&			\left(\begin{array}{cccccccccccc}
			A_{\mathcal{T},\mathcal{F}_u}(\mathcal{L},\mathcal{L}) & \mathbf{0} \\
			A_{\mathcal{T},\mathcal{F}_u}(\overline{\mathcal{L}},\mathcal{L}) & A_{\mathcal{T},\mathcal{F}_u}(\overline{\mathcal{L}},\overline{\mathcal{L}})
		\end{array}
		\right),
	\end{eqnarray*}
	where $ A_{\mathcal{T},\mathcal{F}_u}(\mathcal{L},\overline{\mathcal{L}})$ is a  zero matrix since
	$$j\in\left\{\delta_{m}^{(\ell,\tau)}(i_v,t):m\in\mathcal{R},t\in\mathcal{T},i_v\in\mathcal{F}_u\right\}=\mathcal{L}\subseteq\mathcal{N}$$
	according to \eqref{eqn:AB} and \eqref{eqn:beta_m},
	i.e., $A_{t,i_v}(\delta_{m}^{(\ell,\tau)},j)=0$ when $j\in \overline{\mathcal{L}}$ for any $t\in\mathcal{T},i_v\in\mathcal{F}_u,\delta_{m}^{(\ell,\tau)}\in\mathcal{L},m\in\mathcal{R}$.
	Therefore, $A_{\mathcal{T},\mathcal{F}_u}(\mathcal{L},\mathcal{L})$ is an $(n-d-1)s^{n-d-1}\times (n-d-1)s^{n-d-1}$ invertible matrix. Definitely, by reordering the rows, $A_{\mathcal{T},\mathcal{F}_u}(\mathcal{L},\mathcal{L})$ is equivalent to $B_{\mathcal{R},\mathcal{F}_u}(\mathcal{T})$ following the elementary transformation below,
	\begin{eqnarray*}
		A_{\mathcal{T},\mathcal{F}_u}(\mathcal{L},\mathcal{L})
		&\rightarrow&\left(\begin{array}{cccccccccccc}
			A_{0,\mathcal{F}_u}(\mathcal{L},\mathcal{L})(0)\\
			\vdots   \\
			A_{n-d-2,\mathcal{F}_u}(\mathcal{L},\mathcal{L})(0)\\
			A_{0,\mathcal{F}_u}(\mathcal{L},\mathcal{L})(1)\\
			\vdots   \\
			A_{n-d-2,\mathcal{F}_u}(\mathcal{L},\mathcal{L})(1)\\
			\vdots  \\
			A_{0,\mathcal{F}_u}(\mathcal{L},\mathcal{L})(s^{n-d-1}-1)\\
			\vdots   \\
			A_{n-d-2,\mathcal{F}_u}(\mathcal{L},\mathcal{L})(s^{n-d-1}-1)\\		
		\end{array}
		\right)\\
		&=&\left(\begin{array}{c}
			B_{0,I_u}(0)\\
			\vdots \\
			B_{0,I_u}(n-d-2)\\
			B_{1,I_u}(0)\\
			\vdots \\
			B_{1,I_u}(n-d-2)\\
			\vdots \\
			B_{s^{n-d-1}-1,I_u}(0)\\
			\vdots \\
			B_{s^{n-d-1}-1,I_u}(n-d-2)
		\end{array}
		\right)\\
		&\rightarrow &B_{\mathcal{R},\mathcal{F}_u}(\mathcal{T}),
	\end{eqnarray*}
	where the equation holds from \eqref{eqn:abm}.
	
	This completes the proof.
\end{proof}

\begin{Lemma}\label{lemma:h}
	For any $h$ failed nodes $\mathcal{F}=\{i_0,i_1,\dots,i_{h-1}\}$, the recovery matrix $R$ in \eqref{eqn:repair matrix0} is nonsingular.
\end{Lemma}
\begin{proof}
	We only show the case of node $i_0$ as  that of other $h-1$ nodes is similar.
	For the failed node $i_0$, we  split the submatrix of \eqref{eqn:repair matrix0} into two parts: the first $s=d-k+1$ rows, denoted by $D(\mathcal{S})$ and $B_{m,i_v}(\mathcal{S}),m\in\mathcal{R},v\in[1,n-d)$, and the last $r-s=n-d-1$ rows,
	denoted by $D(\mathcal{S}_r)$ and $B_{m,i_v}(\mathcal{S}_r)$, respectively. Recall from \eqref{zigzag:A_i:parameter} and \eqref{eqn:D} that $D\in \mathbb{F}_q^{r\times s}$ with
	\begin{eqnarray*}
		D = \left(\begin{array}{cc}
			D(\mathcal{S})\\
			D(\mathcal{S}_r)
		\end{array}
		\right)=\left(
		\begin{array}{cccccc}
			1  & 1  & \cdots & 1\\
			1  & 1  & \cdots & \gamma^{i_0+1}\\
			\vdots & \vdots & \ddots & \vdots\\
			1  & \gamma^{i_0+1}  & \cdots & \gamma^{i_0+1}\\
			\gamma^{i_0+1}  & \gamma^{i_0+1}  & \cdots & \gamma^{i_0+1}\\
			\gamma^{i_0+1}  & \gamma^{i_0+1}  & \cdots & \gamma^{2(i_0+1)}\\
			\vdots & \vdots & \vdots & \vdots
		\end{array}
		\right).
	\end{eqnarray*}
	
	Then, we perform the following elementary  row transformation on $R$, i.e.,
	\begin{eqnarray*}
		\resizebox{1\hsize}{!}{$
			R\rightarrow R'=
			\left(\begin{array}{cccccccc}
				D(\mathcal{S}) & \mathbf{0} & \cdots & \mathbf{0} &{B_{0,\mathcal{F}_0}}(\mathcal{S}) \\
				\mathbf{0} & D(\mathcal{S}) &  \cdots & \mathbf{0}  &{B_{1,\mathcal{F}_0}}(\mathcal{S})\\
				\vdots & \vdots &\ddots & \vdots  & \vdots \\
				\mathbf{0} & \mathbf{0} & \cdots & 	D(\mathcal{S} )  &{B_{s^{n-d-1}-1,\mathcal{F}_0}}(\mathcal{S}) \\				
				\mathbf{0} & \mathbf{0} & \cdots & \mathbf{0} &\Lambda'
			\end{array}\right),
			$}
	\end{eqnarray*}
	where $\mathcal{F}_0=\{i_1,i_2,\dots,i_{n-d-1}\}$ and
	\begin{eqnarray*}
		\Lambda'&=& {B_{\mathcal{R},\mathcal{F}_0}}(\mathcal{S}_r)-\gamma^{i_0+1}{B_{\mathcal{R},\mathcal{F}_0}}(\mathcal{T})\\
		&=&
		\left(\begin{array}{cccccccc}
			{B_{0,\mathcal{F}_0}}(\mathcal{S}_r)-\gamma^{i_0+1}{B_{0,\mathcal{F}_0}}(\mathcal{T})\\
			{B_{1,\mathcal{F}_0}}(\mathcal{S}_r)-\gamma^{i_0+1}{B_{1,\mathcal{F}_0}}(\mathcal{T})\\
			\vdots\\
			{B_{s^{n-d-1}-1,\mathcal{F}_0}}(\mathcal{S}_r)-\gamma^{i_0+1}{B_{s^{n-d-1}-1,\mathcal{F}_0}}(\mathcal{T})
		\end{array}\right).
	\end{eqnarray*}
	
	According to \eqref{zigzag:A_i:parameter}-\eqref{A_i^t}, for $t\in\mathcal{T},v\in[1,n-d)$ and $m\in\mathcal{R}$, we know
	$$A_{s+t,i_v}(m)= \gamma^{i_v+1}A_{t,i_v}(m).$$
	Thus,
	$$(A_{s+t,i_v}(\mathcal{L},\mathcal{L}))(m)= \gamma^{i_v+1}(A_{t,i_v}(\mathcal{L},\mathcal{L}))(m).$$
	Following \eqref{eqn:abm}, it means,
	$${B_{m,i_v}}(s+t)= \gamma^{i_v+1}{B_{m,i_v}}(t),$$
	i.e.,
	$$	B_{m,i_v}(\mathcal{S}_r)= \gamma^{i_v+1}{B_{m,i_v}}(\mathcal{T}),$$
	which results in
	\begin{eqnarray*}
		\det(\Lambda')=(\gamma^{i_1+1}-\gamma^{i_0+1})(\gamma^{i_2+1}-\gamma^{i_0+1})\cdots\\ (\gamma^{i_{n-d-1}+1}-\gamma^{i_0+1})\det (B_{\mathcal{R},\mathcal{F}_0}(\mathcal{T})).
		%
		%\left(\begin{array}{cccccccc}
			%	{B_{0,\mathcal{F}_u}}(\mathcal{T})\\
			%		{B_{1,\mathcal{F}_u}}(\mathcal{T}) \\
			%		\vdots  \\
			%		{B_{s^{n-d-1}-1,\mathcal{F}_u}}(\mathcal{T})
			%	\end{array}\right).
	\end{eqnarray*}
	
	Following Lemma \ref{Lemma:B}, we know $\det(\Lambda')\ne 0$ since $\det (B_{\mathcal{R},\mathcal{F}_0}(\mathcal{T}))\ne 0$ and $(\gamma^{i_v+1}-\gamma^{i_0+1})\ne 0,i_v\in\mathcal{F}_0$,
	where $\gamma$ is a primitive element of $\F_q$.
	This implies that  the recovery matrix $R$ given by \eqref{eqn:repair matrix0} is nonsingular, which completes the proof.
\end{proof}

The following result is a direct consequence of  Theorem  \ref{P-invert} and Lemma \ref{lemma:h}.
\begin{Theorem}
	Any $h$ failed nodes of $(n,k,d,h,N)$ MSR codes with sub-packetization $N=(d-k+h)s^n$ generated in Construction \ref{con} can be optimally cooperative repaired, where $s=d-k+1$.
\end{Theorem}

\begin{Example}
	%Continue with Example \ref{Example_1}.
	Let $h=2$ and generate $d-k+h=4$ instances of the $(n=7,k=2,d=4,h=1,N=3^7)$ MSR code $\mathcal{C}'$ over $\mathbb{F}_{2^3}$.
	In this way, we  obtain an $(n=7,k=2,d=4,h=2,N=4\cdot3^7)$ MSR code $\mathcal{C}$.  Denote  the column vector of length $N=4\cdot3^7$ stored at node $i$ by $\mathbf{f}_i=((\mathbf{f}_i^{(0)})^\top,(\mathbf{f}_i^{(1)})^\top, (\mathbf{f}_i^{(2)})^\top,(\mathbf{f}_i^{(3)})^\top)^\top$, where $\mathbf{f}_{i}^{(w)}=\{f_{i,a}^{(w)}:a\in[0,3^7)\}^\top, i\in[0,6], w\in[0,3]$ is a column vector of length $3^7$.
	
	Assume that the nodes $0$ and $1$ fail, and the helper nodes are $j\in[3,6]$, thus, the unconnected node is node $2$.
	Take  the repair of the node $0$ as an example. For brevity, we only indicate the summing and grouping steps and the form of $D$, and the other details are similar to those of Example \ref{Example_2}.
	
	\textbf{Step 1 (Summing):}	
	For $a\in [0,3^7),t\in[0,4]$, by \eqref{eqn:simply0}, we get $3^7$ SPCEs of node $0$ as
	\begin{align*}
		&\zeta_{t,0,a}f_{0,a(0,t)}^{(0)}+\zeta_{t,0,a(0,1)}f_{0,a(0,1+t)}^{(1)}+\zeta_{t,0,a(0,2)}f_{0,a(0,2+t)}^{(2)}\\
		&+\sum_{i=1}^2\zeta_{t,i,a}\left(f_{i,a(i,t)}^{(0)}+f_{i,a(0,1;i,t)}^{(1)}+f_{i,a(0,2;i,t)}^{(2)}\right) \\
		=&	-\sum_{j=3}^6\zeta_{t,j,a}\left(f_{j,a(j,t)}^{(0)}+f_{j,a(0,1;j,t)}^{(1)}+f_{j,a(0,2;j,t)}^{(2)}
		\right).
	\end{align*}

	%following \eqref{eqn:zeta}, \eqref{eqn:A_{t,i,a}} and \eqref{eqn:a-th parity}.	
	
	\textbf{Step 2 (Grouping):}
	According to \eqref{eqn:a}, divide all the symbols $\{a:a\in[0,3^7)\}$ into $81$ groups, i.e.,
	\begin{eqnarray*}
		\mathcal{A}_0&=&\{0,1,2,\dots,25,26\},\\
		\mathcal{A}_1&=&\{27,28,29,\dots,52,53\},\\
		\mathcal{A}_2&=&\{54,55,56,\dots,79,80\},\\
		&\vdots&\\
		\mathcal{A}_{80}&=&\{3^7-27,3^7-26,3^7-25,\dots,3^7-1\}.
	\end{eqnarray*}
	Following \eqref{eqn:b}, partition the $27$ symbols in $\mathcal{A}_0$ to $3$ groups as
	\begin{eqnarray*}
		&&\mathcal{B}_{0,0}=\{0,5,7,11,13,15,19,21,26\},\\
		&&\mathcal{B}_{0,1}=\{1,3,8,9,14,16,20,22,24\},\\
		&&\mathcal{B}_{0,2}=\{2,4,6,10,12,17,18,23,25\}.
	\end{eqnarray*}
	
	Taking the $(\ell=0,\tau=0)$-th group as an example, we can obtain
	\begin{eqnarray*}
		D = \left(\begin{array}{cc}
			D([0,2])\\
			D([3,4])
		\end{array}
		\right)=\left(
		\begin{array}{cccccc}
			1  & 1  & 1\\
			1  & 1  & \gamma\\
			1  & \gamma & \gamma\\
			\gamma& \gamma & \gamma\\
			\gamma& \gamma & \gamma^2
		\end{array}
		\right),
	\end{eqnarray*}
	where the invertibility of $D([0,2])$ is immediate.
\end{Example}

\subsection{Computational complexity}
In this subsection, we analyze the computational complexity of the proposed $(n,k,d,h,N=(d-k+h)s^n)$ 	MSR code $\mathcal{C}$.
According to the Grouping rules in \eqref{eqn:a} and \eqref{eqn:b}, for a fixed failed node $i_u \in \mathcal{F}$, we identify a $$\Big(rN/ \big((d-k+h)s^{d+1}\big)\Big) \times \Big(rN/ \big((d-k+h)s^{d+1}\big)\Big)$$ recovery matrix $R$ such that, for $\ell \in [0, s^d)$ and $\tau \in [0, s=d-k+1)$, we can express the parity check equations in \eqref{eqn:simply0} as $RX=MU$,
where $X$ and $U$ are column vectors of length $rN/ \big((d-k+h)s^{d+1}\big)$ and $dN/ \big((d-k+h)s^{d+1}\big)$ corresponding to the unknowns in the subset $\mathcal{B}_{\ell,\tau} \subseteq [0,s^n)$ of failed and unconnected nodes, and the data downloaded from the $d$ helper nodes, respectively. Therein, $M$ is the $$\Big(rN/ \big((d-k+h)s^{d+1}\big)\Big) \times \Big(dN/ \big((d-k+h)s^{d+1}\big)\Big)$$ coefficient matrix. Solving $X = R^{-1}(M U)$ requires $$r(r+d) \Big(N/ \big((d-k+h)s^{d+1}\big)\Big)^2$$ multiplications and $$r(r+d) \Big(N/ \big((d-k+h)s^{d+1}\big)\Big)^2 - r \Big(N/ \big((d-k+h)s^{d+1}\big)\Big)$$ additions. Consequently, the computational complexity for recovering the failed nodes is $O\big(({N\over d-k+h})^{2}\big)$ when $\ell$ and $\tau$ range over $[0, s^d)$ and $[0, s=d-k+1)$, respectively.

It is noteworthy that, the sub-packetization level of MSR codes is, by necessity, typically of exponential order in this literature \cite{alrabiah2021exponential,balaji2022lower}.
Our constructions are subject to the same fundamental limitation. Consequently, the proposed codes are theoretically meaningful when the sub-packetization level is sufficiently large, and become practically relevant when the code length is relatively small.
%	In addition, although the theory supports that the failed nodes download data from arbitrary $k\le d\le n-h$ helper nodes during the Download phase, smaller $d$ values (e.g., $d=k+1$, $k+2$, corresponding to lower repair bandwidth) represent more typical application scenarios.

\begin{table}[t]
	\begin{center}
		\caption{Some parameters of MSR codes with optimal cooperative repair property over $\mathbb{F}_{2^8}$.}
		\label{tab:paramaters}
		\renewcommand\arraystretch{1.2}
		\begin{tabular}{|c|c|c|c|c|c|c|c|}
			\hline
			$d$& $k$&  $n$ &  $N$   & Ref.\\
			\hline
			\multirow{3}{*}{$k+1$}& \multirow{3}{*}{-} &  $n\le 128$ & $(h+1)2^n$     & \cite{ye2020new} \\	
			\cline{3-5}
			&  &  $n\le 127$ & $(h+1)2^{\lfloor n/2\rfloor}$     & \cite{zhang2024construction}\\	
			\cline{3-5}
			&   & $n\le 255$ & $(h+1)2^n$     & The code $\mathcal{C}$\\	
			\hline
			\multirow{3}{*}{$k+2$} &\multirow{3}{*}{-} &   $n\le 85$ & $(h+2)3^n$     & \cite{ye2020new}\\	
			\cline{3-5}
			&  &  $n\le 85$ & $(h+2)3^{\lfloor n/2\rfloor}$     & \cite{zhang2024construction}\\	
			\cline{3-5}
			&  &  $n\le 255$ & $(h+2)3^n$     & The code $\mathcal{C}$\\	
			\hline
			\multirow{3}{*}{$n-3$} & \multirow{6}{*}{$n-10$} & $n\le 32$ & $(h+7)8^n$     & \cite{ye2020new}\\	
			\cline{3-5}
			&    &$n\le 31$ & $(h+7)8^{\lfloor n/2\rfloor}$     & \cite{zhang2024construction}\\	
			\cline{3-5}
			&  &  $n\le 255$ & $(h+7)8^n$     & The code $\mathcal{C}$\\	
			\cline{1-1}\cline{3-5}
			\multirow{3}{*}{$n-2$} &  &  $n\le 28$ & $(h+8)9^n$     & \cite{ye2020new}\\	
			\cline{3-5}
			&  &  $n\le 28$ & $(h+8)9^{\lfloor n/2\rfloor}$     & \cite{zhang2024construction}\\	
			\cline{3-5}
			&  &  $n\le 255$ & $(h+8)9^n$     & The code $\mathcal{C}$\\	
			\hline
		\end{tabular}
	\end{center}
\end{table}

Recall that supporting a wide range of code parameters under a fixed small finite field is practically important. Specifically,  $\mathbb{F}_{2^8}$ is widely used in practical storage systems as a balance between implementation efficiency and coding flexibility. Under such a small-field setting, the MSR constructions in \cite{ye2020new} and \cite{zhang2024construction} support only a limited range of code lengths, whereas our construction significantly enlarges the admissible range.
As shown in Table~\ref{tab:paramaters}, when $d=n-3$, the maximum code length supported by the schemes in \cite{ye2020new} and \cite{zhang2024construction} is $32$, whereas our construction yields valid codes for all lengths from $33$ to $255$. In addition, the proposed construction supports all codelength with $n<256$ under any parameters $d$ and $k$, while existing methods support at most $n\le128$ under the same setting.
Moreover, the parameters of MSR codes with the optimal repair property are compared in Table \ref{tab:comparison}, which indicates that when restricting the field size to $\mathbb{F}_{2^3}$, the codes in \cite{liu2022new,ye2020new,ye2018cooperative} and \cite{zhang2024construction} do not exist at all.
These improvements in our work enable substantially greater flexibility in practical system design, allowing storage systems to adopt both short and long coding configurations while maintaining the same implementation platform and finite-field arithmetic.

\begin{Remark}
	The Download and Cooperative phases must be executed sequentially in practical implementations, as all symbols required in the cooperative phase are recovered  during the Download phase.
\end{Remark}

\section{Conclusion}\label{sec:conclusion}
In this paper, we proposed an optimal cooperative repair scheme for $(n,k,d,h,N)$ MSR codes, capable of recovering any $h$ failed nodes by downloading data from $k \le d \le n-h$ helper nodes. Notably, the required field size is $\mathbb{F}_q$ with $q \ge n+1$, which is smaller than all known constructions to the best of our knowledge. As a further direction, it is of interest to design MSR codes that achieve the optimal cooperative access property, thereby enabling additional reductions in I/O overhead in practical systems. Moreover, developing MSR codes with optimal cooperative repair property in rack-aware storage models remains an intriguing and challenging open problem.
%	\section*{Appendix}
%		\textbf{The proof of Lemma \ref{lemma:recover}:}

%	\section*{Acknowledgment}
%	
%	The authors would like to thank the Associate Editor,
%	Prof. Mingyue Ji, and the anonymous reviewers, whose
%	comments and suggestions improved the presentation of this
%	paper.
%	

\section{Appendix}
\subsection{Proof of Lemma \ref{lemma:recover}}

First of all,   according to \eqref{eqn:a} and \eqref{eqn:b}, for a given $m\in[0,s^{n-d-1})$ there exists a unique
$$\mathbf{a}=(a_0,\dots,a_{i_0},\dots,a_{i_1},\dots,a_{i_{n-d-1}},\dots,a_{n-1})\in \mathcal{B}_{\ell,\tau}$$
such that
\begin{eqnarray}\label{eqn:m-a}
	(a_{i_0},a_{i_1},\dots,a_{i_{n-d-1}})=(m_0,m_1,\dots,m_{u-1},a_{i_u},m_{u},\nonumber\\m_{u+1},
	\dots,m_{n-d-2}),
\end{eqnarray}
where $a_{i_u},u\in[0,h)$, is determined by \eqref{eqn:b} and the other components of $\mathbf{a}$ are determined by \eqref{eqn:a}.
Thus, hereafter we set $\mathbf{a}=\delta_{m}^{(\ell,\tau)}$ for $m\in[0,s^{n-d-1})$, since \eqref{eqn:a} and \eqref{eqn:b} depend on  $m,\ell$ and $\tau$ following \eqref{eqn:m-a}.

Next, we prove that the LHS of \eqref{eqn:simply1} in matrix form  can be expressed as $R\cdot X$ with
\begin{eqnarray}\label{eqn:X-R}
	X=\left(X_0,...,X_{s^{n-d-1}-1},X_{i_0},\dots,X_{i_{u-1}},X_{i_{u+1}},\dots,\right.\nonumber\\\left.X_{i_{n-d-1}}\right)^{\top},
\end{eqnarray}
where
\begin{eqnarray}\label{eqn:unknows_j}
	X_j=\left(f_{i_u,\delta_{j}^{(\ell,\tau)}}^{(0)},f_{i_u,\delta_{j}^{(\ell,\tau)}(i_u,1)}^{(1)},\dots, f_{i_u,\delta_{j}^{(\ell,\tau)}(i_u,s-2)}^{(s-2)},\right.\nonumber\\
	\left.f_{i_u,\delta_{j}^{(\ell,\tau)}(i_u,s-1) }^{(u+s-1)}\right),\,j\in[0,s^{n-d-1})
\end{eqnarray}
is a row vector of length  $s$  at node $i_u$ and
\begin{eqnarray}\label{eqn:B_i'}
	X_{i_v}	=\left(\sum_{w=0}^{s-2}f_{i_v,\delta_{0}^{(\ell,\tau)}(i_u,w)}^{(w)}+f_{i_v,\delta_{0}^{(\ell,\tau)}(i_u,s-1)}^{(u+s-1)},\dots,\right.\nonumber\\
	\left.\sum_{w=0}^{s-2}f_{i_v,\delta_{s^{n-d-1}-1}^{(\ell,\tau)}(i_u,w)}^{(w)}+f_{i_v,\delta_{s^{n-d-1}-1}^{(\ell,\tau)}(i_u,s-1)}^{(u+s-1)}\right)
\end{eqnarray}
is a row vector of length $s^{n-d-1}$  at node $i_{v},v\in[0,n-d)\backslash\{u\}$.
%, where identity holds as $\{\mathbf{a}+ \langle w-t\rangle  \mathbf{e}_{n,i_u}+  t\mathbf{e}_{n,i_v}:a\in\mathcal{B}_{\ell,\tau}, w\in[0,d-k+h),t\in[0,r)\}=\{\mathbf{a}+ w  \mathbf{e}_{n,i_u}:a\in\mathcal{B}_{\ell,\tau},w\in[0,d-k+h)\}$ following \eqref{eqn:b}.

The equation \eqref{eqn:unknows_j} is obvious from the fact that
\begin{itemize}
	\item [F1.] When $\mathbf{a}=\delta_{m}^{(\ell,\tau)},m\in[0,s^{n-d-1})$, only $X_{m}$  among $X_j,j\in [0,s^{n-d-1})$ appears in LHS of \eqref{eqn:simply1}  for all $t\in[0,r)$.
\end{itemize}

For \eqref{eqn:B_i'}, it is seen from \eqref{eqn:simply1} that for a fixed $t\in[0,r)$, the unknowns of node $i_v\in\mathcal{F}\cup\mathcal{U}\backslash\{i_u\}$ form
\begin{eqnarray}\label{eqn:unknows}
	\resizebox{0.9\linewidth}{!}{
		$\begin{aligned}
			&\left\{\sum_{w=0}^{s-2}f_{i_v,\delta_{0}^{(\ell,\tau)}(i_u,w-t;i_v,t)}^{(w)}+f_{i_v,\delta_{0}^{(\ell,\tau)}(i_u,s-1-t;i_v,t)}^{(u+s-1)},\dots,\right. \\
			&\sum_{w=0}^{s-2}f_{i_v,\delta_{s^{n-d-1}-1}^{(\ell,\tau)}(i_u,w-t;i_v,t)}^{(w)}+\left.f_{i_v,\delta_{s^{n-d-1}-1}^{(\ell,\tau)}(i_u,s-1-t;i_v,t)}^{(u+s-1)}\right\}.
		\end{aligned}
		$}
\end{eqnarray}

Note that if $\delta_{m}^{(\ell,\tau)}\in \mathcal{B}_{\ell,\tau}$, so does $\delta_{m}^{(\ell,\tau)}(i_u,-t;i_v,t)$ by \eqref{eqn:b}. Therefore, all the $r$ ($t\in[0,r)$) sets in \eqref{eqn:unknows} are  equivalent to
\begin{eqnarray*}
	\left\{\sum_{w=0}^{s-2}f_{i_v,\delta_{0}^{(\ell,\tau)}(i_u,w)}^{(w)}+f_{i_v,\delta_{0}^{(\ell,\tau)}(i_u,s-1)}^{(u+s-1)},\dots,\right.\hspace{1cm}\\ \left.\sum_{w=0}^{s-2}f_{i_v,\delta_{s^{n-d-1}-1}^{(\ell,\tau)}(i_u,w)}^{(w)}+f_{i_v,\delta_{s^{n-d-1}-1}^{(\ell,\tau)}(i_u,s-1)}^{(u+s-1)}\right\}
\end{eqnarray*}
regardless of the order,  which is essentially formed by  the elements in \eqref{eqn:B_i'}.

Precisely, we observe the following fact that
\begin{itemize}
	\item [F2.]  When  $\mathbf{a}=\delta_{m}^{(\ell,\tau)},m\in[0,s^{n-d-1})$, by \eqref{eqn:m-a}, only the $j$-th component of $X_{i_v}$   appears in LHS of \eqref{eqn:simply1}  for given  $t\in[0,r)$, with
	\begin{eqnarray*}\label{eqn:m'}
		j=\left\{\begin{array}{lll}
			m(v,t), & \text{~~if~}v<u,\\
			m(v-1,t),& \text{~~if~} v >u.
		\end{array}
		\right.
	\end{eqnarray*}
\end{itemize}

Now, we are ready to show the details of  matrix $R$. Let us consider the $r$ SPCEs in subgroup \eqref{eqn:simply1} for given $\mathbf{a}=\delta_{m}^{(\ell,\tau)}\in\mathcal{B}_{\ell,\tau},m\in[0,s^{n-d-1})$, which corresponds to the block matrix $R_{m}$ in the  $m$-th block row of $(r s^{n-d-1})\times (rs^{n-d-1})$ matrix $R$. To be consistent with \eqref{eqn:X-R}, we express $R_{m}$ in matrix
form as
\begin{eqnarray*}\label{eqn:R-X}
	R_{m}=\left(D_{m,0},\dots, D_{m,s^{n-d-1}-1}, B_{m,i_0},\dots,B_{m,i_{u-1}},\right.\\
	\left.B_{m,i_{u+1}},\dots,B_{m,i_{n-d-1}}\right),
\end{eqnarray*}
where $D_{m,i}$ and $B_{m,i_v}$ are $r\times s$ block matrices and $r\times s^{n-d-1}$ block matrices, respectively.

%To make notation simply, we denote $(\delta_{m}^{(\ell,\tau)})$ as $\beta$ in the following matrix.
According to F1, $D_{m,j}$ is a zero matrix except for
$j=m$, then we have $D_{m,m}$ as
\begin{equation}\label{eqn:matrixi_u}\setlength{\arraycolsep}{1.2pt}
	\resizebox{.95\hsize}{!}{$
		\begin{split}
			D_{m,m}=&	\left(\begin{array}{cccccc}
				\zeta_{0,i_u,\delta_{m}^{(\ell,\tau)}} & \zeta_{0,i_u,\delta_{m}^{(\ell,\tau)}(i_u,1)} & \cdots &\zeta_{0,i_u,\delta_{m}^{(\ell,\tau)}(i_u,s-1)}   \\
				\zeta_{1,i_u,\delta_{m}^{(\ell,\tau)}(i_u,-1)} & \zeta_{1,i_u, \delta_{m}^{(\ell,\tau)}} & \cdots &\zeta_{1,i_u,\delta_{m}^{(\ell,\tau)}(i_u,s-2) }  \\
				\vdots& \vdots & \ddots & \vdots\\
				\zeta_{r-1,i_u,\delta_{m}^{(\ell,\tau)}(i_u,1-r) } & \zeta_{r-1,i_u,\delta_{m}^{(\ell,\tau)}(i_u,2-r)} & \cdots &\zeta_{r-1,i_u,\delta_{m}^{(\ell,\tau)}(i_u,s-r) }
			\end{array}\right)\\
			=&\left(\begin{array}{cccccc}
				1 & 1 & \cdots & 1  \\
				\lambda_{i_u,\langle (\delta_{m}^{(\ell,\tau)})_{i_u}-1\rangle} & \lambda_{i_u, (\delta_{m}^{(\ell,\tau)})_{i_u}} & \cdots &\lambda_{i_u,\langle (\delta_{m}^{(\ell,\tau)})_{i_u} +s-2\rangle}  \\
				\vdots& \vdots & \ddots & \vdots\\
				\prod\limits_{j\in[0,r-2]}\lambda_{i_u,\langle(\delta_{m}^{(\ell,\tau)})_{i_u}+j+1-r\rangle} &
				\prod\limits_{j\in[0,r-2]}\lambda_{i_u,\langle (\delta_{m}^{(\ell,\tau)})_{i_u}+j+2-r\rangle} & \cdots &\prod\limits_{j\in[0,r-2]}\lambda_{i_u,\langle (\delta_{m}^{(\ell,\tau)})_{i_u} +j+s-r\rangle}
			\end{array}\right),
		\end{split}
		$}\end{equation}
where the equation holds due to \eqref{eqn:zeta}. As $\langle  (\delta_{m}^{(\ell,\tau)})_{i_u}+i \rangle$ ranges over $[0,s)$ with
$i$ varying from $-1$ to $s-2$, the coefficient matrix \eqref{eqn:D} can be obtained from \eqref{eqn:matrixi_u} by columns exchanging.

For every $m\in[0,s^{n-d-1}),v\in [0,n-d)\backslash\{u\}$, the submatrix $B_{m,i_v}$ corresponds to the coefficients of  unknowns $X_{i_v}$. Thus, the equation \eqref{eqn:BIJ} follows from  F2. This completes the proof.

%When $a=\delta_{m}^{(\ell,\tau)},m\in[0,s^{h-1})$ enumerates $\mathcal{B}_{\ell,\tau},\ell \in[0,s^d), \tau\in[0,s)$,
%since
%\begin{eqnarray*}
%	&&	\{f_{i_v,\mathbf{a}+ \langle w-t\rangle  \mathbf{e}_{n,i_u}+  t\mathbf{e}_{n,i_v}}^{(w)}:w\in[0,s-2],t\in[0,r)\}\cup	\{f_{i_v,\mathbf{a}-  \langle t+1\rangle \mathbf{e}_{n,i_u}+ t\mathbf{e}_{n,i_v}}^{(u+s-1)}:u\in[0,h),t\in[0,r)\}
%	\nonumber\\
%	&	=&\{f_{i_v,\mathbf{a}}^{(w)}:w\in [0,s-2]\cup\{u+s-1\},u\in[0,h)\},
%\end{eqnarray*}
%we obtain the combination of all symbols of node $i_v\in\mathcal{F}\backslash\{i_u\}$ in $X_{i_v},v\in[0,h)]\backslash\{u\}$.

%\subsection{Proof of Lemma \ref{Lemma:B}}

\bibliographystyle{IEEEtranS}
\bibliography{lyjBib}

Han Cai (Member, IEEE) received the B.S. and M.S. degrees in mathematics
from Hubei University, Wuhan, China, in 2009 and 2013, respectively, and the
Ph.D. degree from the Department of Communication Engineering, Southwest
Jiaotong University, Chengdu, China, in 2017. From October 2015 to 2017,
he was a Visiting Ph.D. Student with the Faculty of Engineering, Information
and Systems, University of Tsukuba, Japan. From 2018 to 2021, he was a
Post-Doctoral Fellow at the School of Electrical and Computer Engineering,
Ben-Gurion University of the Negev, Israel. In 2021, he joined Southwest
Jiaotong University, where he currently hold a tenure-track position. His
research interests include coding theory and sequence design.

Xiaohu Tang (Fellow, IEEE) received the B.S. degree in applied mathematics
from Northwest Polytechnic University, Xi’an, China, in 1992, the M.S.
degree in applied mathematics from Sichuan University, Chengdu, China, in
1995, and the Ph.D. degree in electronic engineering from Southwest Jiaotong
University, Chengdu, in 2001.
From 2003 to 2004, he was a Research Associate with the Department of
Electrical and Electronic Engineering, The Hong Kong University of Science
and Technology. From 2007 to 2008, he was a Visiting Professor with the
University of Ulm, Germany. Since 2001, he has been with the School of
Information Science and Technology, Southwest Jiaotong University, where he
is currently a Professor. His research interests include coding theory, network
security, distributed storage, and information processing for big data.
Dr. Tang was a recipient of the National Excellent Doctoral Dissertation
Award in 2003, China, the Humboldt Research Fellowship in 2007, Germany,
and the Outstanding Young Scientist Award by NSFC, China, in 2013.
He served as an Associate Editor for several journals, including IEEE
TRANSACTIONS ON INFORMATION THEORY and IEICE Transactions on
Fundamentals, and served on a number of technical program committees of
conferences.

\end{document}